\documentclass{aa}
\usepackage{txfonts}
\usepackage[dvips]{color}
\usepackage{graphicx}
\usepackage{natbib}
\bibpunct{(}{)}{;}{a}{}{,}
\usepackage{captcont}
\usepackage{subfigure}
\usepackage{sidecap}

\begin{document}

\title{Gas density drops inside dust cavities of transitional disks around young stars observed with ALMA}
\titlerunning{Gas in transitional disks}
\author{N. van der Marel\inst{1} 
\and E.F. van Dishoeck\inst{1,2}
\and S. Bruderer\inst{2}
\and L. P{\' e}rez\inst{3}\thanks{Jansky Fellow of the National Radio Astronomy Observatory}
\and A. Isella\inst{4}
}
\institute{Leiden Observatory, Leiden University, P.O. Box 9513, 2300 RA Leiden, the Netherlands
\and Max-Planck-Institut f\"{u}r Extraterrestrische Physik, Giessenbachstrasse 1, 85748 Garching, Germany
\and National Radio Astronomy Observatory, P.O. Box O, Socorro, NM 87801, USA
\and Rice University, 6100 Main Street, Houston, TX 77005, USA
}
\date{Accepted by A\&A, April 2015}

\abstract{Transitional disks with large dust cavities are important laboratories to study planet formation and disk evolution. Cold gas may still be present inside these cavities, but the quantification of this gas is challenging. The gas content is important to constrain the origin of the dust cavity.}
  {We use Atacama Large Millimeter/submillimeter Array (ALMA) observations of $^{12}$CO 6--5 and 690 GHz (Band 9) continuum of five well-studied transitional disks. In addition, we analyze previously published Band 7 observations of a disk in $^{12}$CO 3--2 line and 345 GHz continuum. The observations are used to set constraints on the gas and dust surface density profiles, in particular the drop $\delta_{\rm gas}$ of the gas density inside the dust cavity.}
{The physical-chemical modeling code DALI is used to analyze the gas and dust images simultaneously. We model SR21, HD135344B, LkCa15, SR24S and RXJ1615-3255 (Band 9) and J1604-2130 (Band 7). The SED and continuum visibility curve constrain the dust surface density. Subsequently, the same model is used to calculate the $^{12}$CO emission, which is compared with the observations through spectra and intensity cuts. The amount of gas inside the cavity is quantified by varying the $\delta_{\rm gas}$ parameter.}
{Model fits to the dust and gas indicate that gas is still
present inside the dust cavity for all disks but at a reduced level. The gas surface density drops inside the cavity by at least a factor 10, whereas the dust density drops by at least a factor 1000. Disk masses are comparable with previous estimates from the literature, cavity radii are found to be smaller than in the 345 GHz SubMillimeter Array (SMA) data. }
{The derived gas surface density profiles suggest clearing of the cavity by one or more companions in all cases, trapping the millimeter-sized dust at the edge of the cavity.} 


\maketitle

\section{Introduction}
Planets are formed in disks of dust and gas that surround
young stars \citep[e.g.][]{WilliamsCieza2011}. Of particular interest are the so-called transitional disks with cleared out cavities in the inner part of the dust disk. These
disks are expected to be in the middle of active evolution and possibly planet formation. Transitional disks were originally identified through a dip in the mid infrared part of the Spectral Energy Distribution (SED), indicating a lack of small micron-sized dust grains close to the star. They were modeled as axisymmetric dust disks with inner cavities \citep[e.g.][]{Strom1989,Brown2007,Espaillat2014} subsequently confirmed by millimeter interferometry \citep[e.g.][]{Dutrey2008, Brown2009, Andrews2011,Isella2012}. In recent years it has become clear that the gas and dust do not necessarily follow the same radial distribution in transition disks \citep[e.g.][]{Pontoppidan2008}. High accretion rates ($\sim10^{-8}$ M$_{\odot}$ yr$^{-1}$, e.g. \citealt{Calvet2005,Najita2007}) indicate that the dust cavities are not empty, but gas is still flowing through the gap. Near infrared observations of CO rovibrational lines indicate the presence of warm gas inside the dust cavity \citep{Salyk2007,Pontoppidan2008,Salyk2009,Brittain2009,Brown2012a}. However, these data do not trace  the bulk of the molecular gas. ALMA CO pure rotational line observations reveal directly the presence of gas inside the dust cavity \citep{vanderMarel2013,Bruderer2014,Perez2014,Casassus2013,Zhang2014}. Also the size distribution of the dust is possibly not uniform throughout the disk: micron- and millimeter-sized grains may have different distributions \citep{Garufi2013,vanderMarel2013} and millimeter- and centimeter-sized grains can show radial variations \citep{Perez2012}.

Quantifying the density structure of the gas inside the dust hole is essential to distinguish between the possible clearing mechanisms of the dust in the inner part of disk: photoevaporation \citep{Clarke2001}, grain growth \citep{DullemondDominik2005},  clearing by (sub)stellar or planetary companions \citep[e.g.][]{LinPapaloizou1979}. Also, instabilities at the edges of dead zones have been involved to millimeter-dust rings and asymmetries \citep[e.g.][]{Regaly2012}. Planet candidates have been found in cavities of the transition disks T Cha \citep{Huelamo2011}, LkCa~15 \citep{KrausIreland2012} and HD~100546 \citep{Quanz2014}, indicating planet clearing as a possible scenario. 

Since detection of planets in disks is challenging, an indicator of the clearing by a (planetary) companion can be inferred from the different structure of the gas compared to the dust in several of the observed disks as shown by planet-disk interaction models in combination with dust evolution models \citep[e.g.][]{Pinilla2012b}. Companion clearing changes the gas surface density profile from that of a full disk: the density at and around the orbital radius of the planet is decreased while at the outer edge of this gas gap the density slightly increases. Due to the induced pressure gradients along the sides of this pressure bump, dust particles will drift towards the edge of the gap, get trapped and grow efficiently to larger sizes. This results in a dust ring of millimeter-sized dust or radial dust trap \citep{Pinilla2012b}. The decrease in gas density due to companion clearing depends on the mass of the companion and other parameters such as the viscosity \citep{Zhu2011,Dodson-RobinsonSalyk2011,Pinilla2012b,Mulders2013,Fung2014}. Other effects that play a role in dynamic clearing are for example migration and planet accretion luminosity \citep[e.g.][]{KleyNelson2012}. Therefore, the amount of gas inside a cavity can help to constrain the mass of the companion in the planet disk clearing and dust trapping scenario if the viscosity is known. In comparison, the surface density of the millimeter-sized dust drops significantly deeper inside the cavity  compared to the gas due to the trapping of millimeter-sized dust \citep{Pinilla2012b}. Therefore, the quantification of gas \emph{and} dust inside the cavity indicates if the trapping mechanism due to a planet is at work. Dust trapping at the edges of dead zones will also result in millimeter-sized dust rings, but in that scenario the gas surface density remains the same as for a full disk, apart from a minor density increase (less than 50\%) at the edge \citep{Lyra2015}.

The rovibrational lines used in previous work of transitional disks can indicate the presence of CO inside the dust cavity and the inner radius, but are not suitable for quantification of the total gas density. The emission is very sensitive to the gas temperature, which is high but uncertain in the models in this regime. 
Also, the dust opacity of the inner disk, which provides shielding, is uncertain. The pure rotational CO lines provide better constraints on the gas surface density inside the cavity. The brightest molecular lines in the submillimeter regime are the rotational $^{12}$CO lines. Before ALMA, interferometric data generally did not have the sensitivity and spatial resolution to detect rotational $^{12}$CO emission inside the cavity. However, even if CO line emission is detected, this emission can not be directly translated into densities, especially inside the dust cavity: the CO abundance is not constant throughout the disk with respect to molecular hydrogen due to photodissociation by the UV radiation from the star and freeze out onto dust grains in colder regions close to the midplane \citep{Zadelhoff2001,Aikawa2002}. Furthermore, the gas temperature is decoupled from the dust temperature in the surface of the disk, particularly at the cavity wall that is directly heated and inside the dust-depleted cavity \citep[e.g.][]{KampDullemond2004,Jonkheid2004,GortiHollenbach2008}. 

To determine the amount of gas that is present inside a dust cavity, the physical and chemical structure of the gas needs to be modeled. For this we make use of the physical-chemical modeling tool DALI \citep{Bruderer2012,Bruderer2013}, which solves the heating-cooling balance of the gas and chemistry simultaneously to determine the gas temperature, molecular abundances and molecular excitation for a given density structure. Due to the disk Keplerian rotation and gas thermal and turbulent motions, the $^{12}$CO emission along
any line of sight that intercepts the disk is a spectral line comprising of an optically thick central part and potentially optically thin wings. With spatially resolved CO observations it is possible to infer emission from different lines of sight, in contrast to previous single dish spectral analysis. \citet{Bruderer2014} demonstrated for $^{12}$CO 6--5 lines in Oph IRS 48 that the emission in the optically thin wings can be used to constrain the gas density. In this case the observations are found to be consistent with a physical disk model with two gas density drops inside the millimeter dust cavity. However, Oph IRS 48 has a very low disk mass (constrained by C$^{17}$O observations) so even the $^{12}$CO line is close to optically thin. For more massive disks, the bulk of the $^{12}$CO emission may remain optically thick even if the gas density inside the dust depleted cavities drops by one or two orders of magnitude. In that case,
the $^{12}$CO emission originates from deeper in the disk (closer to the disk midplane) where the gas is colder then in the surface, causing a drop in the optically thick line intensity \citep{Bruderer2013}. 

\begin{table*}[!ht]\small
\caption{Stellar properties}
\label{tbl:stellar}
\begin{tabular}{llllllllll}
\hline
\hline
Target&SpT&$L_*$&$M_*$&$R_*$&$T_{\rm eff}$&$\dot{M}$&$d$&$A_V$&Ref.\\
&&($L_{\odot}$)&($M_{\odot}$)&($R_{\odot}$)&(K)&($M_{\odot}$ s$^{-1}$)&(pc)&(mag)&\\
\hline
SR21&G3&10&1.0&3.2&5830&$<1\cdot10^{-9}$&120&6.3&1,2,3\\ 
HD135344B&F4&7.8&1.6&2.2&6590&$6\cdot10^{-9}$&140&0.3&4,5,6\\
LkCa15&K3&1.2&1.0&1.7&4730&$2\cdot10^{-9}$&140&1.5&6,7,8\\
RXJ1615-3255&K5&1.3&1.1&2.0&4400&$4\cdot10^{-10}$&185&0.4&5,9\\
SR24S&K2&2.5&0.7&2.7&4170&$7\cdot10^{-8}$&120&7.0&5,10,11\\ 
J1604-2130&K2&0.76&1.0&1.4&4550&$1\cdot10^{-11}$&145&0.66&12-14\\   
\hline
\end{tabular}
\\
\begin{footnotesize}
1) \citet{Dunkin1997}, 2) \citet{Pontoppidan2008}, 3) \citet{Grady2009}, 4) \citet{Prato2003}, 5) \citet{Andrews2011}, 6) \citet{Espaillat2010}, 7) \citet{Pietu2007}, 8) \citet{Ingleby2009}, 9) \citet{Wichmann1997}, 10) \citet{Natta2006}, 11) \citet{Luhman1999}, 12) \citet{Dahm2008}, 13) \citet{Mathews2012}, 14) \citep{Carpenter2014}
\end{footnotesize}
\end{table*}

To date, a total of 9 transitional disks with large resolved millimeter dust cavities has been observed in ALMA Cycle 0 in dust continuum and $^{12}$CO. In this paper, we present observations of the disks SR21, HD135344B, RXJ1615-3255, LkCa15 and SR24S through $^{12}$CO 6--5 observations \citep{Perez2014} and J1604-2130 through $^{12}$CO 3--2 observations \citep{Carpenter2014,Zhang2014} . With DALI we determine the density structure of the gas and dust that is consistent with the SED, dust continuum interferometry and the $^{12}$CO observations. If available, information on the hot gas and dust from the literature is included. The goals of this study are to determine the amount of the gas and dust inside the dust hole and the implications for potentially embedded planets, assuming that the millimeter dust rings are indeed dust traps. The derived physical models in this study can be used as a starting point for future ALMA observations of CO isotopologues, whose emission is optically thin and which therefore directly trace the density structure. Our analysis of the gas density structure is similar to that used for Oph IRS 48 through $^{12}$CO 6--5 and C$^{17}$O 6--5 line observations as presented in \citet{Bruderer2014}. The analysis of HD142527, which has been observed in $^{12}$CO, $^{13}$CO and C$^{18}$O 6--5, 3--2 and 2--1, is left for future studies. \citet{SPerez2014} have modeled the $J$=2--1 CO isotopologues and determined the gas mass remaining inside the cavity of that disk. Other transition disks that have been observed in ALMA Cycle 0 are HD 100546 \citep{Walsh2014}, TW Hya \citep{Rosenfeld2012}, Sz 91 \citep{Canovas2015} and MWC 758 (not yet published), but since their dust holes remain unresolved in the current data they are not considered in this study. 

The paper is structured as follows. In Section 2.1 we describe the details of the ALMA observations. In Section 2.2 we present moment maps of the $^{12}$CO observations and the derived stellar position and orientation. The modeling approach and constraints from the literature are presented in Section 3. Section 4 presents the modeling results. Section 5 discusses the implications for embedded planets in the disk. 

\section{Data}
\begin{figure*}[!ht]
\begin{center}
\subfigure{\includegraphics[scale=1,trim=0 10 0 10]{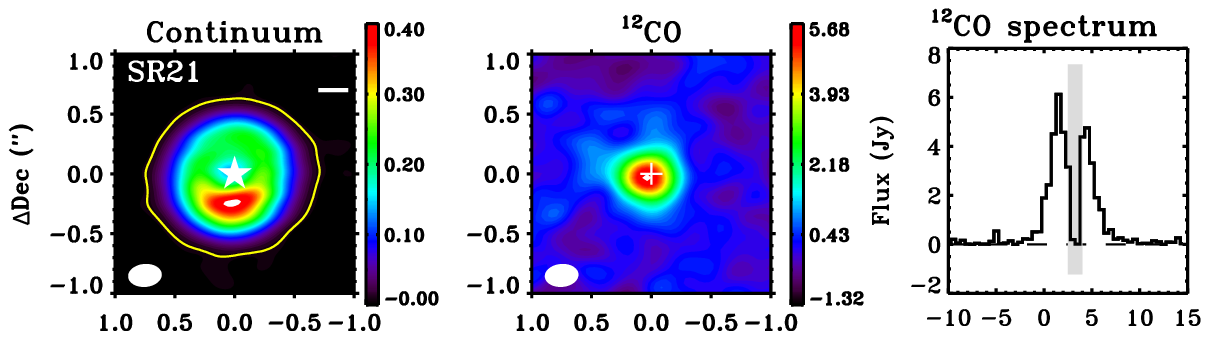}}\\
\subfigure{\includegraphics[scale=1,trim=0 10 0 10]{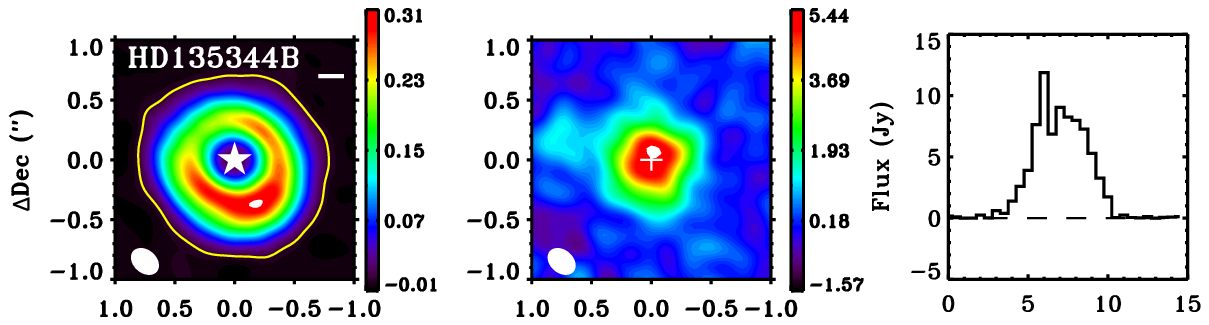}}\\
\subfigure{\includegraphics[scale=1,trim=0 10 0 10]{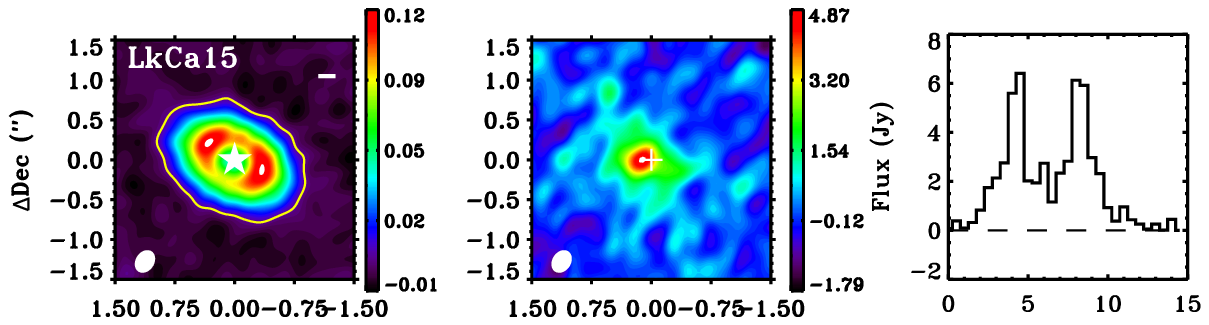}}\\
\subfigure{\includegraphics[scale=1,trim=0 10 0 10]{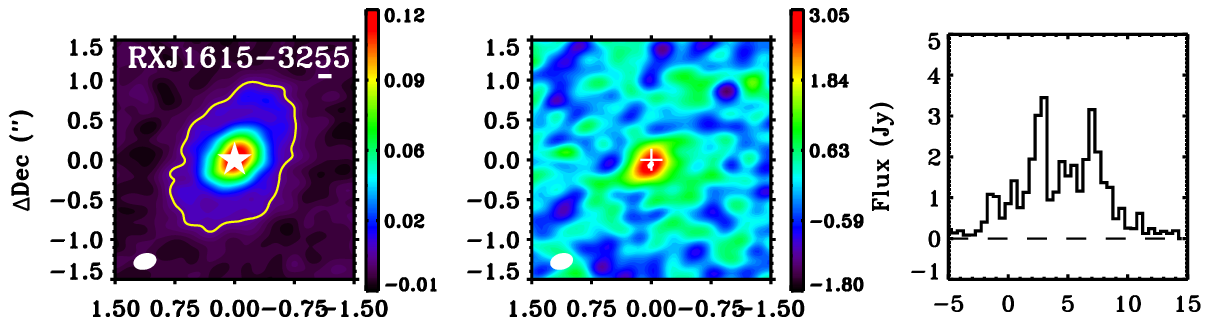}}\\
\subfigure{\includegraphics[scale=1,trim=0 10 0 10]{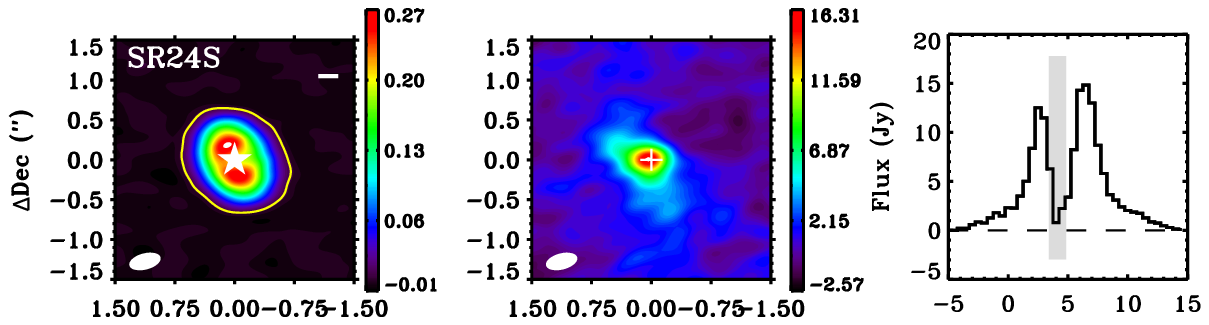}}\\
\subfigure{\includegraphics[scale=1,trim=0 10 0 10]{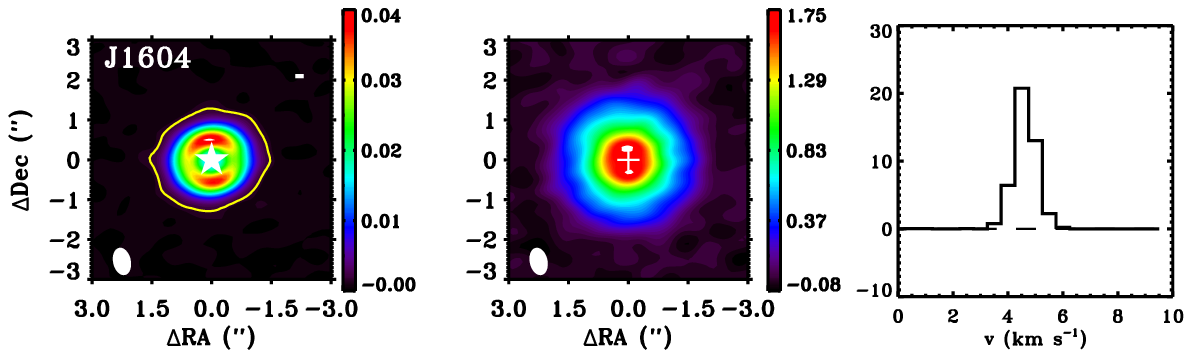}}\\
\end{center}
\caption{ALMA observations of the continuum and $^{12}$CO line. The first 5 disks show the 690 GHz (440 $\mu$m) continuum and the $^{12}$CO $J$=6--5 line, the 6th is the 345 GHz (880 $\mu$m) continuum and $^{12}$CO $J$=3--2 line. {\bf Left:} Continuum image. The stellar position is indicated by a white star, the white bar in the upper right  corner indicates the 30 AU scale and the yellow contour gives the 3$\sigma$ detection limit. The colorbar units are given in Jy beam$^{-1}$. ; {\bf Center:} zero-moment $^{12}$CO map. The colorbar units are given in Jy beam$^{-1}$ km s$^{-1}$; {\bf Right:} $^{12}$CO spectrum integrated over the entire disk. The dashed line indicates the zero flux level and the grey areas indicate the parts of the spectrum affected by foreground absorption (seen in SR21 and SR24S). The beam is indicated in each map by a white ellipse in the lower left corner.}
\label{fig:intensity}
\end{figure*}

\begin{table*}[!ht]\small
\caption{Properties of the ALMA observations}
\label{tbl:obsproperties}
\begin{tabular}{llllllllllll}
\hline
\hline
Target&Derived position&Line&Beam&Beam&rms$_{\rm line}$\tablefootmark{a}&Peak$_{\rm line}$\tablefootmark{a}&rms$_{\rm cont}$&Peak$_{\rm cont}$&PA&$i$&$\varv_{\rm src}$\\
&(J2000)&&size (")&PA ($^{\circ}$)&(mJy)&(mJy)&(mJy)&(mJy)&($^{\circ}$)&($^{\circ}$)&(km/s)\\
\hline
SR21&16:27:10.27 -24:19:13.03&$^{12}$CO 6--5&0.28$\times$0.19&-85&\phantom{5}65&1660&1&409&14&15&2.75\\
HD135344B&15:15:48.43 -37:09:16.36&$^{12}$CO 6--5&0.26$\times$0.19&49&114&2010&1.8&318&63&20&7.05\\
LkCa15&04:39:17.80 +22:21:03.21&$^{12}$CO 6--5&0.30$\times$0.23&-33&120&1202&1.8&118&60&55&6.1\\
RXJ1615-3255&16:15:20.23 -32:55:05.40&$^{12}$CO 6--5&0.29$\times$0.20&-74&102&960&1.7&123&153&45&4.6\\
SR24S&16:26:58.51 -24:45:37.14&$^{12}$CO 6--5&0.40$\times$0.20&-75&\phantom{5}96&2196&1.9&278&20&45&4.75\\
J1604-2130&16:04:21.65 -21:30:28.90&$^{12}$CO 3--2&0.69$\times$0.44&-78&100&1500&0.20&\phantom{5}38&80&10&4.7\\
\hline
\end{tabular}
\tablefoot{
\tablefoottext{a}{Measured in 0.5 km s$^{-1}$ bins.}
}
\end{table*}

The continuum and $^{12}$CO line observations were obtained during ALMA Cycle 0 between June 2012 and January 2013, with baselines ranging from 20 to 500 m, probing scales from 0.2 to 4 arcseconds. The sources and their properties are summarized in Table \ref{tbl:obsproperties}.

\subsection{Observational details}

Five transitional disks were observed in ALMA program 2011.0.00724.S (PI P{\' e}rez), two of these disks were presented in \citet{Perez2014}. The disks were observed in Band 9 (690 GHz, 440 $\mu$m), with four adjacent spectral windows of 1875 MHz, each with 3840 channels of 488 kHz (0.2 km s$^{-1}$) width, for a total continuum bandwidth of 7.5 GHz. One spectral window was centered on the $^{12}$CO 6--5 line (691.47308 GHz).  The bandpass was calibrated using 3C279, Titan was used as flux calibrator and the phase and amplitude were calibrated using J1427-4206 (HD135344B), J1625-2527 (SR21, SR24S and RXJ1615-3255) and J0510+180 (LkCa15). Each target was observed for $\sim$25 minutes on source. Given the high signal to noise ratio of the continuum, amplitude and phase self-calibration were performed on all data and applied afterwards to the $^{12}$CO data as well. 

Disk J1604-2130 was observed as part of ALMA program 2011.0.00526.S (PI Carpenter), in Band 7 (345 GHz, 880 $\mu$m), with one spectral window centered on the $^{12}$CO 3--2 line (345.79599 GHz) with 488 kHz (0.42 km s$^{-1}$) channel width. The data and observational details are presented in \citet{Zhang2014}.  Due to the longer wavelength, the beam size of these data is considerably larger than that of the Band 9 data.

The data were imaged using CASA v4.2, adopting the Briggs weighting scheme with robust of 0.5. The resulting rms and beam sizes of the line data are given in Table \ref{tbl:obsproperties}. During the cleaning process of the SR24S data, it became clear that there is extended emission that could not be recovered by the shortest baselines of these ALMA observations (50 k$\lambda$, or 4'' scales). 

\subsection{Continuum and line maps}
Figure \ref{fig:intensity} shows the ALMA continuum image, zero-moment $^{12}$CO map and $^{12}$CO spectrum of each of our targets. The spectrum was extracted from the region of the zero-moment map size. The contours indicate the 3$\sigma$ detection limit in each image. The continuum images reveal ring-like structures for all targets except RXJ1615-3255, even though its SED and the 345 GHz visibility curve show signs of a dust cavity \citep{Andrews2011}. The rings clearly show azimuthal asymmetries, which have been confirmed to be real intensity asymmetries for SR21 and HD135344B \citep{Perez2014}. For SR24S and LkCa15 the asymmetries are likely caused by the geometry of the disk ($i\sim45^{\circ}$) and the dust continuum becoming optically thick at this wavelength (0.45 mm). 
The $S/N$ on the continuum varies between $\sim$65 and 400, much higher than previous SubMillimeter Array (SMA) observations.  

The zero-moment $^{12}$CO maps reveal the presence of gas inside the dust hole for all targets, demonstrating different distributions of gas and dust. This does not mean that there can be no decrease in the gas density compared to a full disk density profile, because the $^{12}$CO emission is likely optically thick.  The maps of HD135344B and J1604-2130 show a decrement of emission around the stellar position, directly hinting at a strong decrease of gas density. The size of this gas `hole' is much smaller than the dust hole and comparable to the beam size. For LkCa15 there appears to be a slight offset between the continuum and integrated CO emission peaks; the maps are currently aligned to the same central position. This may be related to the proposed eccentric inner disk \citep{Thalmann2014}, casting a more pronounced shadow on one side of the disk. However, considering the limited significance of this shift, modeling this asymmetry is beyond the scope of this paper.

All $^{12}$CO spectra are consistent with a rotating disk at an inclination $>$15$^{\circ}$, revealing a double-peaked velocity profile, except J1604-2130, which is known to be almost face-on \citep{Mathews2012,Zhang2014}. The spectra of SR21 and SR24S are likely affected by foreground absorption, commonly observed in the Ophiuchus star forming region \citep[e.g.][]{vanKempen2009}, although no single dish spectra of these targets are available to confirm this. The SR21 spectrum is somewhat asymmetric: the red side is narrower than the blue side. Furthermore, the data of SR24S are affected by extended emission outside the field of view of the ALMA observations. This may be related to the suggested circumbinary disk around SR24N, which is located 5" north, and SR24S \citep{Mayama2010}, but this can not be confirmed with the available data. The apparent double peak in the spectrum is caused by foreground absorption: the bulk of this emission is not originating from the disk. 
Therefore, we do not aim to model the gas surface density of this target and only model the dust continuum structure. 

The $^{12}$CO channel maps (binned to 1 km s$^{-1}$) and first moment maps are presented in Figure \ref{fig:channelmap}. We derive the stellar position, inclination, position angle and source velocity from the channel maps and first moment map and overlay the corresponding Keplerian velocity pattern in white contours. The derived parameters are consistent within 5$^{\circ}$ with values from the literature for resolved gas and dust, where available \citep[e.g.][]{Pontoppidan2008,Andrews2011}. For the stellar mass, we use values from the literature (see Table \ref{tbl:stellar}). The derived parameters are given in Table \ref{tbl:obsproperties}. The peak $S/N$ of the channel maps is at least 10.

The channel maps confirm the presence of gas inside the dust cavities. The velocity range that falls within the derived dust cavity radius $r_{\rm cav}$ is given in Table \ref{tbl:velocitycav}. The velocity is calculated using $v_{\rm obs} = \sqrt{\frac{GM_*}{r_{\rm cav}}}\sin{i}$ with $G$ the gravitational constant. The channel maps also reveal that the detected CO emission is more extended in the outer disk than the dust continuum for all sources except SR21, although the signal to noise and image quality are low in the outer parts of the disks. The emission is unlikely to be affected by spatial filtering, as shown in the visibilities (Figure \ref{fig:gasresults}): the emission is well covered down to $\sim$40k$\lambda$, corresponding to 5'' ($>$600 AU) diameter scales. Whether or not the difference in extent is significant needs to be quantified with a physical model, because the dust emission may simply be too weak in the outer disk to be detected while $^{12}$CO remains detectable because it is optically thick. The red-shifted side of both SR21 and HD135344B is somewhat weaker than the blue-shifted side, as also seen in the spectrum (Fig. \ref{fig:intensity}). The location of the decrease is cospatial with the peak of the asymmetry in the continuum map. This is possibly related to a higher continuum optical depth or lower temperature at the location of the asymmetry,  as was also seen for H$_2$CO emission of Oph IRS 48 \citep{vanderMarel2014}. 

\section{Method}
\subsection{Modeling}
Our aim is to constrain a physical disk model that describes both the dust and gas structure of each disk. The main goal is to determine the drop in gas surface density $\delta_{\rm gas}$ inside the dust hole, within the constraints of a model that fits the gas and dust in the outer disk (Fig. \ref{fig:genericmodel}). Therefore the other disk parameters need to be constrained first, as they have a significant influence on the strength of the CO emission inside the cavity: the disk mass relates to the optical depth of the $^{12}$CO lines; the vertical structure affects the temperature by shadowing and the amount of exposed disk surface atmosphere by stellar light; the amount of dust inside the cavity, including the inner disk, shields the CO from photodissociation but also provides cooling; and the location of the cavity wall determines the temperature of the wall, where the bulk of the $^{12}$CO emission is expected to originate (see 3.3.3 for details). To this end, information is combined from the SED, the ALMA continuum and integrated CO visibilities, and the CO disk integrated spectrum and zero-moment map. 

Although the model is not expected to be unique, it sets a structure for the disk that can be tested with future ALMA observations of CO isotopologues. The main goal of this study is to compare the maximum possible decrease of the gas surface density with the minimum decrease of the dust density inside the cavity. If the drop in gas density is less deep than for the millimeter-sized dust, this is direct evidence for the trapping scenario.

The axisymmetric physical-chemical model DALI is used for the analysis \citep[][details next section]{Bruderer2013} , thus any azimuthal asymmetries are ignored. The most prominent asymmetries are those seen in the millimeter-sized dust continuum in SR21 and HD135344B \citep{Perez2014}, but the influence of the millimeter dust on the gas temperature and UV shielding is negligible so axisymmetry can be safely assumed. The small dust grains are the main heating agents of the gas. In the visibilities we therefore use the amplitude ($\sqrt(Real^2+Imag^2)$), because the asymmetries and thus the variations in the imaginary part are small. The visibilities constrain the radial structure of the emission. Due to limited $u,v$-coverage in these Cycle 0 data the visibilities give better constraints on the structure than the images alone, especially on the cavity radius through the location of the first null. 

\subsection{DALI}
The determination of the gas
disk from the $^{12}$CO emission requires a thermo-chemical disk
model, in which the heating--cooling balance of the gas and chemistry
are solved simultaneously to determine the gas temperature and
molecular abundances at each position in the disk. Moreover, even though
the densities in disks are high, the excitation of the rotational
levels may not be in local thermodynamic equilibrium (LTE), and the gas and dust temperature are decoupled, especially inside and at the cavity edge. The DALI model \citep{Bruderer2012,Bruderer2013} uses a combination of a stellar radiation field with a disk density distribution as input. 
DALI solves for the dust temperatures
through continuum radiative transfer from UV to millimeter
wavelengths and calculates the chemical abundances, the molecular
excitation, and the thermal balance of the gas. It was developed for
the analysis of gas emission structures in protoplanetary disks including
transitional disks with varying gas to dust ratios \citep{Bruderer2013}.

DALI uses a reaction network described in detail in
\citet{Bruderer2012} and \citet{Bruderer2013}. It is based on a
subset of the UMIST 2006 gas-phase network
\citep{Woodall2007}. About 110 species and 1500 reactions are
included. In addition to the gas-phase reactions, some basic
grain-surface reactions (freeze-out, thermal and nonthermal
sublimation and hydrogenation like g:O $\to$ g:OH $\to$ g:H$_2$O and
H$_2$/CH$^+$ formation on PAHs \citep{Jonkheid2006} are included. The g:X notation refers to atoms and molecules on the grain surface. The photodissociation
rates are obtained from the wavelength-dependent cross-sections by
\citet{vanDishoeck2006}. The adopted cosmic-ray ionization rate is
$\zeta_{\rm H_2}=5\times 10^{-17}$ s$^{-1}$. X-ray ionization (X-ray luminosity of 10$^{30}$ erg s$^{-1}$) and the effect of
vibrationally exited H$_2$ are also included in the network. Elemental abundances are taken from \citet{Jonkheid2006}. 

\subsection{Approach}
\subsubsection{Physical model}
\begin{figure}
\includegraphics[scale=0.4]{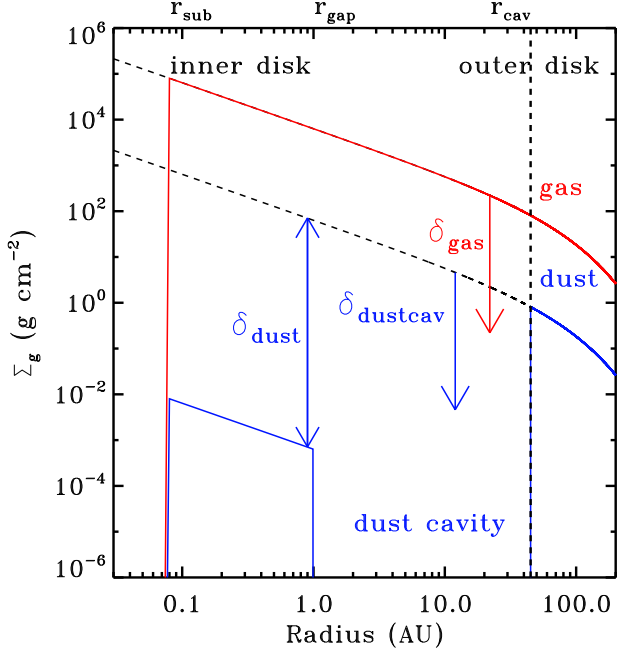}
\caption{Generic surface density profile for the gas and dust.}
\label{fig:genericmodel}
\end{figure}

As a starting point for our models we adopt the physical structure suggested by \citet{Andrews2011}, as implemented by \citet{Bruderer2013} and applied for a similar analysis of Oph IRS 48 \citep{Bruderer2014}. The gas surface density is assumed to be an exponential power-law following the time-dependent viscosity disk model $\nu \sim R^{\gamma}$ \citep{LyndenPringle1974,Hartmann1998}
\begin{equation}
\Sigma(r) = \Sigma_c \left(\frac{r}{r_c}\right)^{-\gamma} {\rm exp}\left(-\left(\frac{r}{r_c}\right)^{2-\gamma}\right)
\end{equation}
and is defined by surface density $\Sigma_c$ at critical radius $r_c$. The power-law index $\gamma$ is taken as 1, in line with the results for normal disks with continuous dust distributions and other transition disk studies \citep{Andrews2011}. The previous gas study of $^{12}$CO observations in IRS~48 \citep{Bruderer2014} demonstrated that the slope of the surface density profile could not be determined with their data, whereas the data in this new study have even lower S/N, so $\gamma$ is not varied.
The outer radius is set to $r_{\rm out}$. The dust density is taken to be $\Sigma(r)$/GDR, with GDR the gas-to-dust-ratio set to 100 initially. 
The inner dust disk is defined by the $\Sigma(r)$ profile, starting from the sublimation radius $r_{\rm sub}=0.07(L_*/L_{\odot})^{1/2}$ (assuming $T_{\rm sub}$=1500 K) out to $r_{\rm gap}$=1 AU, scaled down by a factor $\delta_{\rm dust}$ to fit the near infrared part of the SED. The size of the inner disk is arbitrary as it does not have a significant effect on the near infrared excess.  Between $r_{\rm gap}$ and $r_{\rm cav}$, the cavity radius, the dust density is scaled down by a factor $\delta_{\rm dustcav}$ (the drop in dust density inside the cavity), while the gas density $\Sigma(r)$ is scaled down by a factor $\delta_{\rm gas}$. Figure \ref{fig:genericmodel} shows the generic density structure and Table \ref{tbl:parameters} explains each parameter.

\begin{table*}\small
\begin{center}
\caption{Parameters fitting procedure}
\label{tbl:parameters}
\begin{tabular}{llll}
\hline
\hline
Category&Parameter&Value&Description\\
\hline
Surface density profile ($\Sigma(r)$)&$r_c$&{\it free}&Critical radius\\
&$\Sigma_c$&{\it free}&Gas density at $r_c$\\
&$\gamma$&1&Surface density gradient\\
&GDR&100&Gas to dust ratio\\
\hline
Radial structure&$r_{\rm sub}$&{\it fixed by formula}&Sublimation radius\\
&$r_{\rm gap}$&{\it free}&Size inner disk\\
&$r_{\rm cav}$&{\it free}&Cavity radius\\
&$r_{\rm out}$&{\it free}&Outer radius\\
\hline
Scaling $\Sigma(r)$&$\delta_{\rm dust}$&{\it free}&Dust density drop inner disk ($r_{\rm sub}$--$r_{\rm gap}$)\\
&$\delta_{\rm dustcav}$&{\it free}&Dust density drop inside cavity ($r_{\rm gap}$--$r_{\rm cav}$)\\
&$\delta_{\rm gas}$&{\it free}&Gas density drop inside cavity ($r_{\rm sub}$--$r_{\rm cav}$)\\
\hline
Vertical structure ($h(r)$)&$h_c$&{\it free}&Scale height angle at $r_c$\\
&$\psi$&{\it free}&Flaring angle\\
\hline
Dust properties&sg-pop&0.005--1 $\mu$m&Small dust grain population\\
&lg-pop&0.005 $\mu$m--1 mm&Large dust grain population\\
&$f_{\rm ls}$&{\it free}&Fraction large grains\\
&$\chi$&0.2&Scale height large grain fraction\\
&PAH&0.1\%&PAH fraction w.r.t. ISM abundance\\
\hline
Ionization&$\zeta_{\rm H_2}$&5$\times10^{-17}$s$^{-1}$&Cosmic ray ionization rate\\
&X-ray&10$^{30}$ erg s$^{-1}$&X-ray ionization rate\\
\hline
\end{tabular}
\end{center}
\end{table*}

For the stellar photosphere a blackbody with the luminosity and temperature as in Table \ref{tbl:stellar} is generated. Additional UV excess from accretion is added through a $T_{\rm acc}$=10 000 K blackbody for the $L_{\rm acc}$ corresponding to the measured accretion rate (Table \ref{tbl:stellar}), using 
\begin{equation}
L_{\rm acc} (\nu) = \pi B_{\rm \nu}(T_{\rm acc},\nu) \frac{GM_*}{R_*} \dot{M} \frac{1}{\sigma T_{\rm acc}^4}
\end{equation}
with $B_{\rm \nu}(T_{\rm acc},\nu)$ the Planck blackbody function, $G$ the gravitational constant, $M_*$ and $R_*$ the mass and radius of the star, $\dot{M}$ the mass accretion rate and $\sigma$ the Stefan-Boltzmann constant. For SR21, no accretion luminosity was added since only an upper limit for the accretion is measured. Note that observed values of $\dot{M}$ are uncertain up to an order of magnitude and often variable with time \citep[e.g.][]{Salyk2013}. The interstellar radiation field incident on the disk surface is also included but does not affect the results presented here. 

The vertical structure follows a Gaussian distribution, defined by a scale height angle $h=h_c\left(\frac{r}{r_c}\right)^{\psi}$. A disk in hydrostatic equilibrium will have $\psi$=0.25, while a flat ($h/r$ = constant) disk will have $\psi$ = 0. The near- and far-infrared excess in the SED relate directly with the resulting scale height $h(r)$ at $r_{\rm sub}$ and $r_{\rm cav}$, respectively. Our vertical structure is a simplification and does not account for hydrostatic equilibrium of gas or dust temperature. One of the main effects of the hydrostatic equilibrium is a puffed up inner rim. Previous studies \citep[e.g.][]{Andrews2011} often use a parametrized puffed up inner rim to fit the near- and far-infrared excess by adding a narrow scale height peak, but during our fitting procedure it became clear that a parametrized optically thick inner disk rim blocks a large amount of the UV irradiation, lowering the gas temperature significantly, as also demonstrated in \citet{Woitke2009}.

For the dust opacities we use the same approach as \citet{Andrews2011}, consisting of a small (0.005--1 $\mu$m) and large (0.005 $\mu$m -- 1 mm)  population of dust with a mass fraction $f_{\rm ls}$ in large grains. Settling is parametrized by defining the scale-height of the large grains as $\chi{h}$, i.e. $h$ is reduced by a factor $\chi$ (see \citet{Andrews2011} for definitions). We initially fix $f_{\rm ls}$ and $\chi$ to 0.85 and 0.2 respectively, as adopted by \citet{Andrews2011}. We include PAHs since many of our observed sources show PAH features \citep{Brown2007,Geers2006} but at a low abundance of 0.1\% of the ISM abundance. The turbulent width is set to 0.2 km s$^{-1}$. Stellar parameters taken from the literature are given in Table \ref{tbl:stellar}.

\subsubsection{Model fitting approach}
In the fitting procedure, we use an approach of manual fitting by eye, varying the parameters in a logical order, converging to a model that fits the data, similar to the approach in \citet{Bruderer2014}. We do not use a $\chi^2$ or Markov-Chain-Monte-Carlo (MCMC) method, as the computational time of the models is too long and the number of parameters too large. Although this approach reduces computational time, deriving uncertainties of model parameters, verifying the uniqueness of the fit, and estimating the correlation between parameters is not possible. The fitting parameters are defined in Table \ref{tbl:parameters}.
\begin{enumerate}
\item Fit $\Sigma_c$ and $\delta_{\rm dust}$ roughly through the SED and continuum image, using the $r_{\rm cav}$ value from \citet{Andrews2011}.
\item Adjust $r_{\rm sub}$, $r_{\rm gap}$ and $\delta_{\rm dust}$ if necessary for the NIR excess.
\item Vary the scale height angle to fit the NIR and/or FIR excess in the SED, by changing $h_c$ and $\psi$. 
\item Vary $r_{\rm cav}$ and $r_c$ to match the null and slope of the submillimeter dust visibility curve.
\item If the MIR/FIR excess is too high, adjust the size distribution by increasing $f_{\rm ls}$ and $\chi$.
\item Find the highest possible value for $\delta_{\rm dustcav}$ (drop of dust density inside the cavity) that is still consistent with the dust continuum. 
\item Check the fit to the CO intensity map, spectrum and visibilities assuming the gas-to-dust ratio (GDR) to be 100 in the outer disk. If necessary, truncate the outer radius to fit the spectrum. 
\item Explore small variations in the vertical parameters to change the temperature structure, required to fit the optically thick $^{12}$CO emission.
\item Finally, keep all other parameters constant while varying $\delta_{\rm gas}$ to find the lowest possible value for $\delta_{\rm gas}$ that is still consistent with the CO data. 
\end{enumerate}

We stress that this approach does not result in an unique model, due to the degeneracies in parameter choices and due to $^{12}$CO being optically thick. However, we consider this approach to be reasonable considering the different data sets taken into account, where ALMA is the crucial one. \citet{Carmona2014} also demonstrated a modeling procedure combining different datasets. The submillimeter continuum sets firm constraints on the total flux, the cavity size (the null in the visibility) and the dust extent (through $r_c$) of the disk, as also shown in \citet{Andrews2011}. The mid infrared part of the SED and the $^{12}$CO are both sensitive to the combination of temperature and density structure of the disk, rather than just the density, as both of these emission regimes are mostly optically thick. The temperature is controlled by the adopted vertical structure in the disk, as this determines the direct irradiation by the star, especially at the cavity wall. Therefore, several iterations were required to find a compromise for a decent fit of both the SED and $^{12}$CO. 

\subsubsection{CO emission inside the cavity}
As the CO emission inside the cavity only partly depends on density, the derived $\delta_{\rm gas}$ parameter is merely a constraint within the derived physical model. As discussed above, there are several effects that influence the strength of the emission: most important are gas temperature and CO abundance which are very sensitive to the small dust grains inside the cavity. Different effects can play a role, as discussed extensively in \citet{Bruderer2013}. 
\begin{itemize}
\item Although the bulk of the $^{12}$CO emission is optically thick, the drop in gas surface density can still be constrained to within an order of magnitude with spatially resolved observations. Optically thick emission will slightly drop with decreasing $\delta_{\rm gas}$ because the $\tau=1$ surface moves deeper into the disk, so the emission traces colder regions in the disk, as demonstrated in Section 5.2 and Fig. 9-10 in \citet{Bruderer2013}.
\item An inner disk will cast a shadow on the gas inside the cavity, depending on the amount of dust and scale height. This will lower the gas temperature inside the cavity. On the other hand, the inner disk shields the stellar UV radiation that would photodissociate CO, so the CO abundance is increased. The structure of the inner disk is not well constrained from the available data, while it may have significant effects on the heating of the outer disk.
\item The presence of small dust grains and PAHs inside the cavity shields stellar UV radiation and decreases the photodissociation, thus increasing the CO abundance. The dust contributes to the heating and cooling of the gas, depending on the density. For high gas density, the gas temperature decreases when dust is added due to cooling by gas-grain collisions. For low gas density, the gas temperature decouples and becomes higher than the dust temperature due to photoelectric heating on grains and PAHs.
\item The increase of the gas temperature (increase of small dust grains) or an increase of the UV radiation (decrease of small dust grains) will also lead to an increase of vibrationally pumped H$_2$, again increasing the formation of CO.
\item The presence of small dust grains and PAHs increases the formation of CH$^+$ on the grain surfaces, which leads to increased formation of CO.
\end{itemize}
The consequences for the CO emission of adding small dust, either inside the entire cavity or just in the inner disk, strongly depends on the details of the structure of the disk. This is the reason why also the SED and submillimeter visibilities are included in the modeling procedure, and why a full physical-chemical model is required for analysis of the CO emission.

\begin{figure*}[!ht]
\begin{center}
\subfigure{\includegraphics[scale=0.5,trim=0 20 0 20]{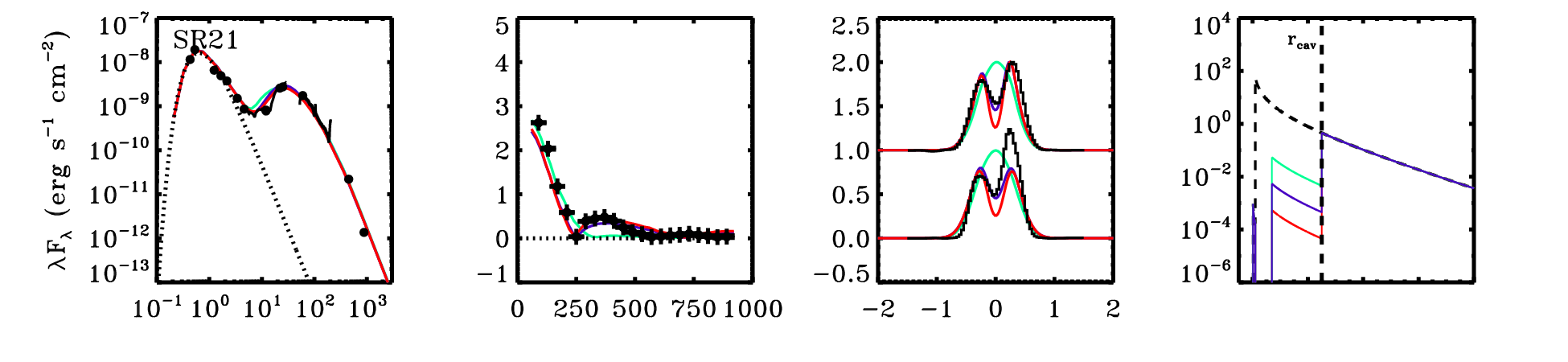}}\\
\subfigure{\includegraphics[scale=0.5,trim=0 20 0 20]{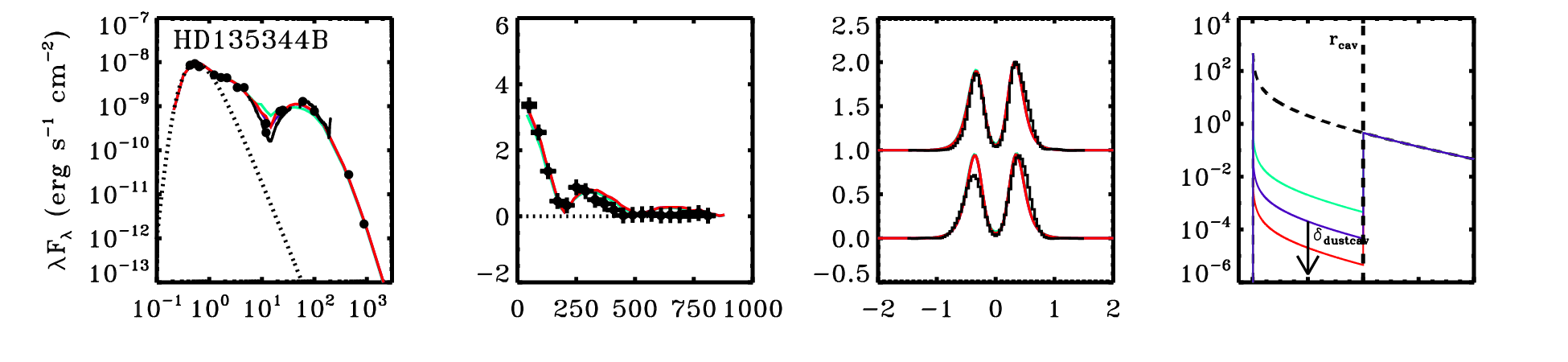}}\\
\subfigure{\includegraphics[scale=0.5,trim=0 20 0 20]{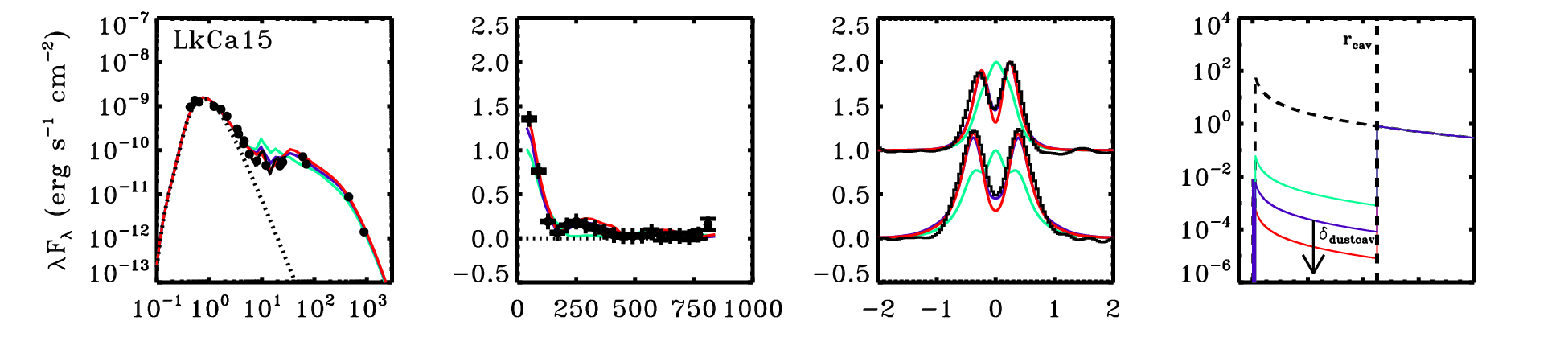}}\\
\subfigure{\includegraphics[scale=0.5,trim=0 20 0 20]{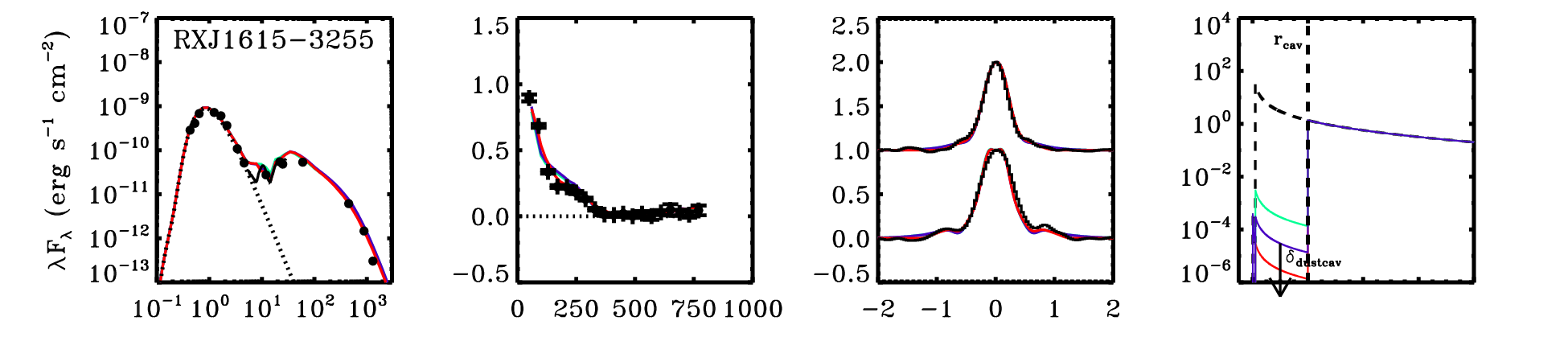}}\\
\subfigure{\includegraphics[scale=0.5,trim=0 20 0 20]{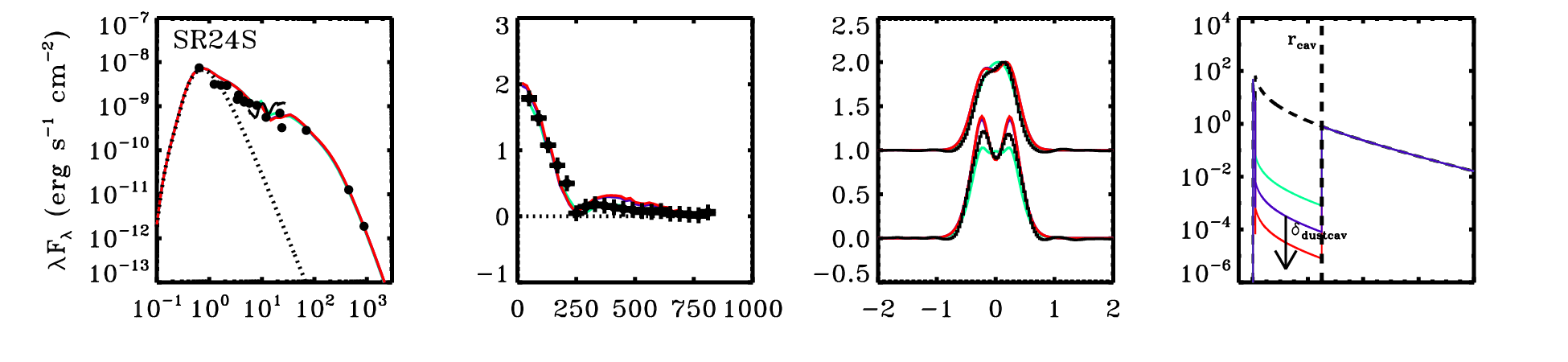}}\\
\subfigure{\includegraphics[scale=0.5,trim=0 20 0 20]{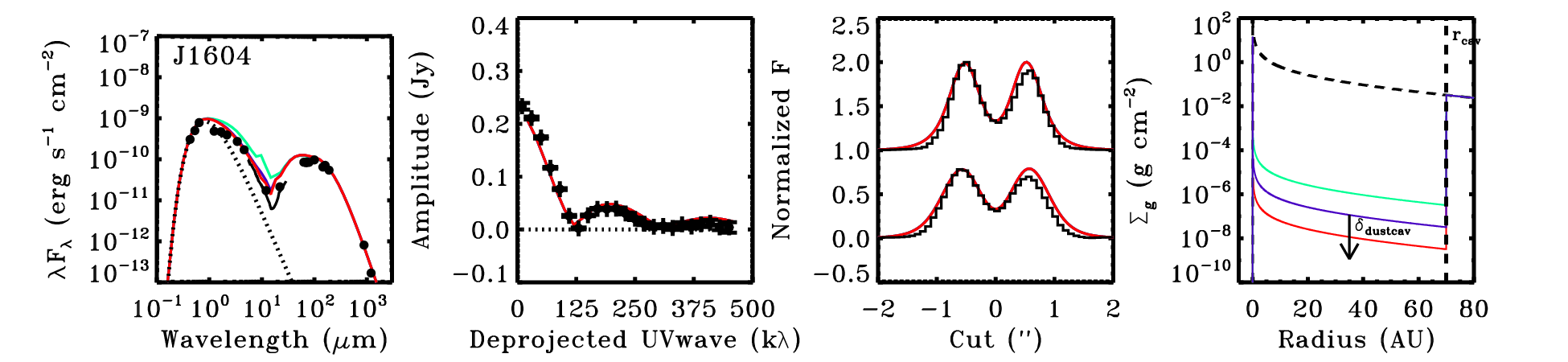}}\\
\end{center}
\caption{Modeling results and observations of the dust surface density, comparing $\delta_{\rm dustcav}$ ranging between 10$^{-2}$ and 10$^{-8}$ as indicated in the right panel. The observations are plotted in black. {\bf Left:} Spectral Energy Distribution; {\bf middle left:} Amplitude of the 690 GHz continuum visibility for the deprojected baselines. The null line is indicated with a dashed line; {\bf middle right:} Normalized intensity cuts through the major (bottom) and minor (top) axis of the 690 GHz continuum image. The model images are convolved with the same beam as the ALMA observations; {\bf right:} The dust surface density profile. Indicated are the $\delta_{\rm dust}$, the drop in density to fit the inner disk through the near infrared emission, and $\delta_{\rm dustcav}$, the minimum drop in dust density inside the cavity to fit the observations. In SR21 (top panel) the region inside 7 AU is assumed to be empty due to additional constraints (see main text). In all other disks $\delta_{\rm dustcav}$ is found to be at most 10$^{-4}$. All other disk parameters are as listed in Table \ref{tbl:fitting}.}
\label{fig:dustresults}
\end{figure*}

\begin{figure*}[!ht]
\begin{center}
\subfigure{\includegraphics[scale=0.5,trim=0 20 0 20]{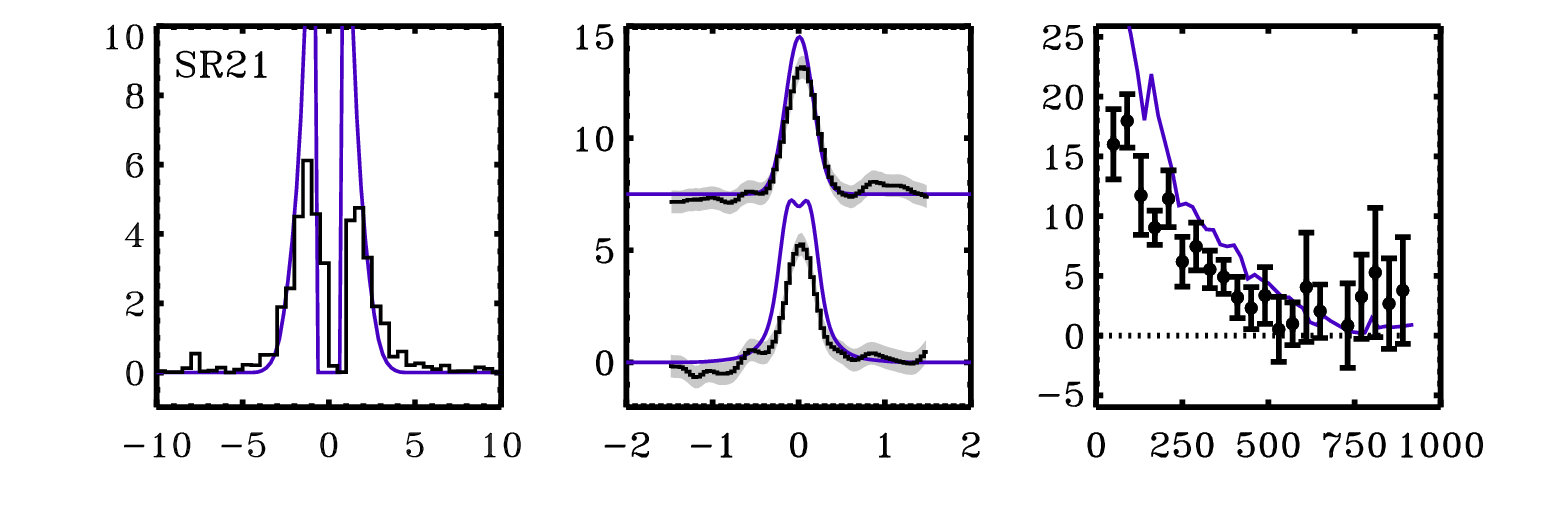}}\\
\subfigure{\includegraphics[scale=0.5,trim=0 20 0 20]{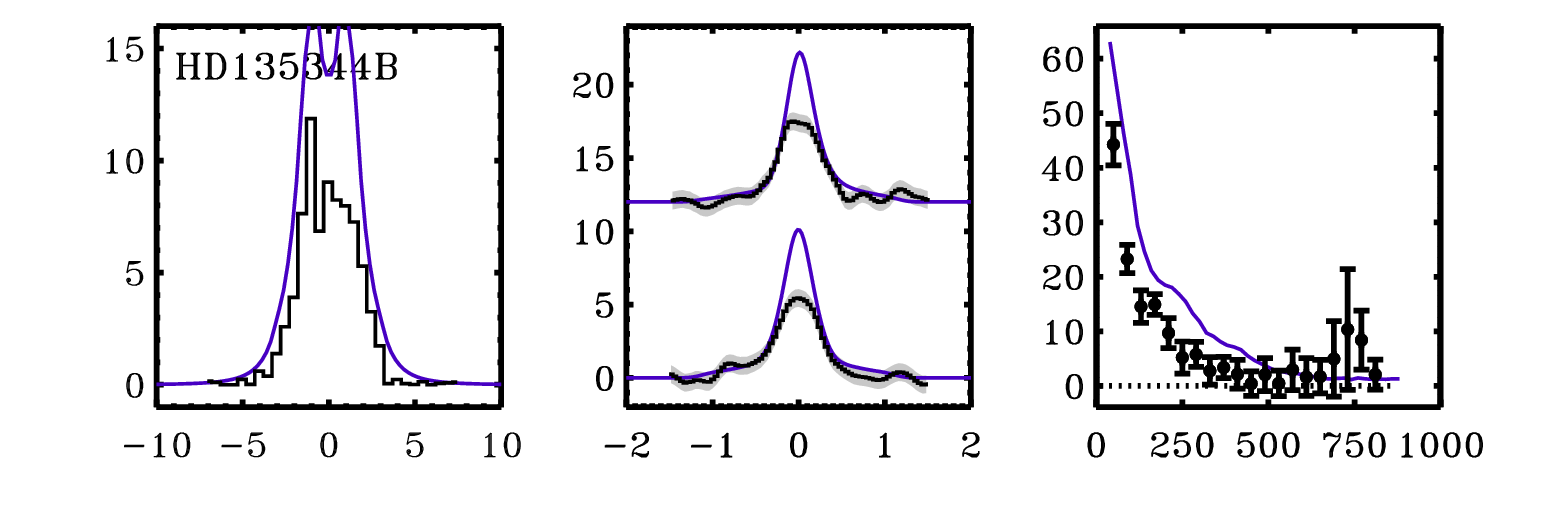}}\\
\subfigure{\includegraphics[scale=0.5,trim=0 20 0 20]{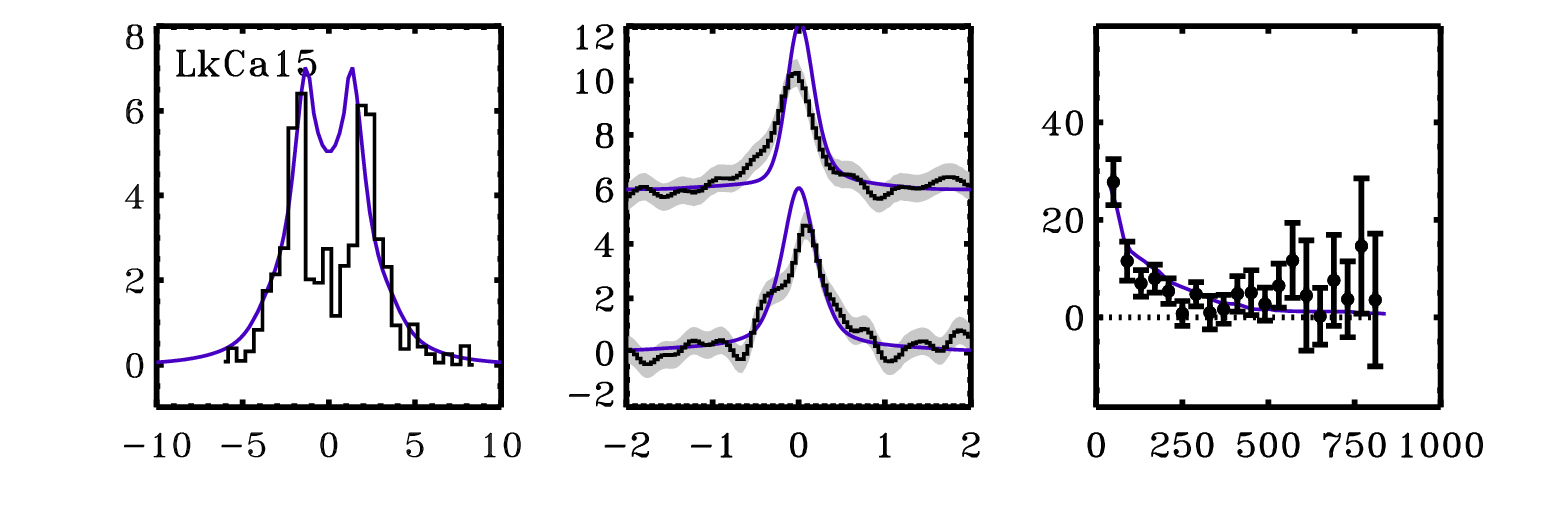}}\\
\subfigure{\includegraphics[scale=0.5,trim=0 20 0 20]{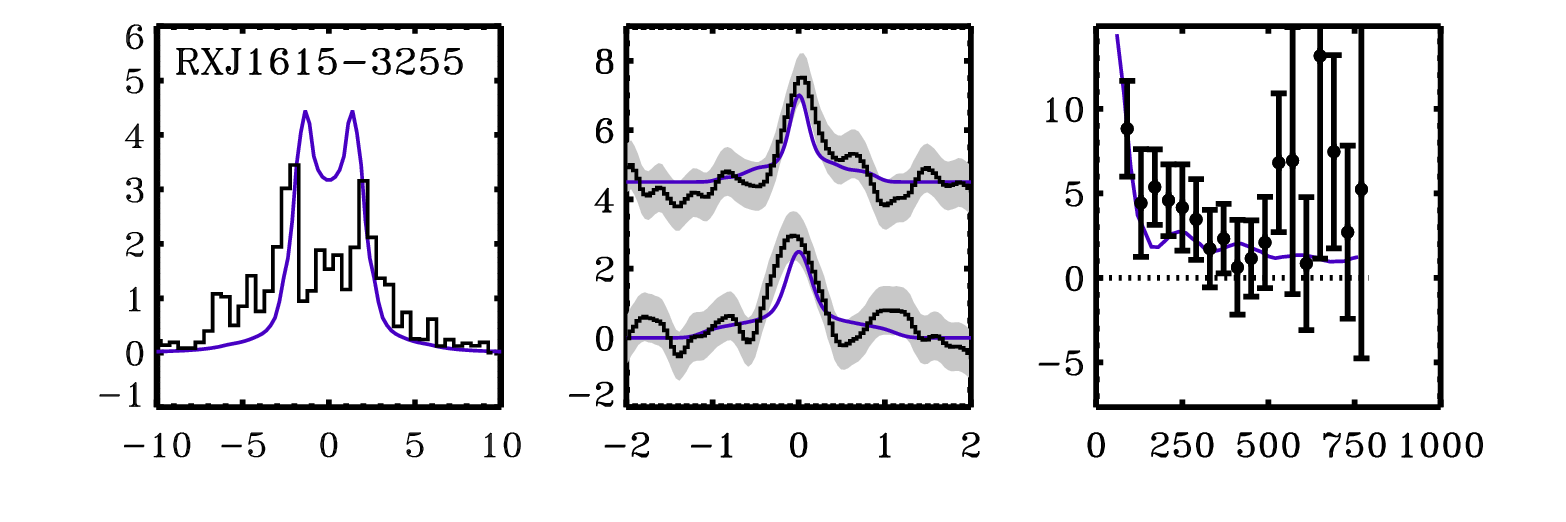}}\\
\subfigure{\includegraphics[scale=0.5,trim=0 20 0 20]{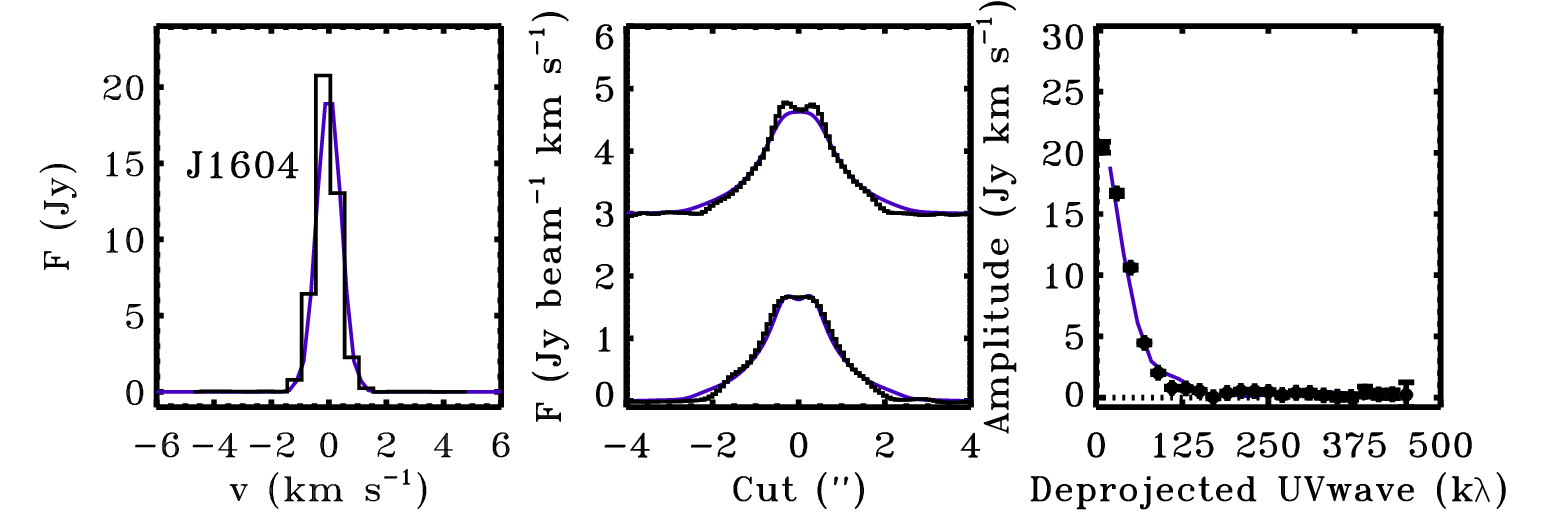}}\\
\end{center}
\caption{Modeling results and observations of the $^{12}$CO emission for the best fitting physical model derived from the dust (see Figure \ref{fig:dustresults} and Table \ref{tbl:fitting}), with $\delta_{\rm gas}$=1 (no drop in gas surface density inside the cavity). The observations are black, the models in purple. {\bf Left:} $^{12}$CO spectrum integrated over the entire disk; {\bf Center:} Intensity cuts through the major (bottom) and minor (top) axis of the $^{12}$CO zero moment map. The noise level is indicated by the gray zone. The model images are convolved with the same beam as the ALMA observations. ; {\bf Right:} Amplitude of the integrated $^{12}$CO visibility for the deprojected baselines. The null line is indicated with a dashed line and error bars indicate the noise.}
\label{fig:gasresults}
\end{figure*}

\section{Results}
\begin{figure*}[!ht]
\begin{center}
\includegraphics[scale=0.5,trim=0 0 0 0]{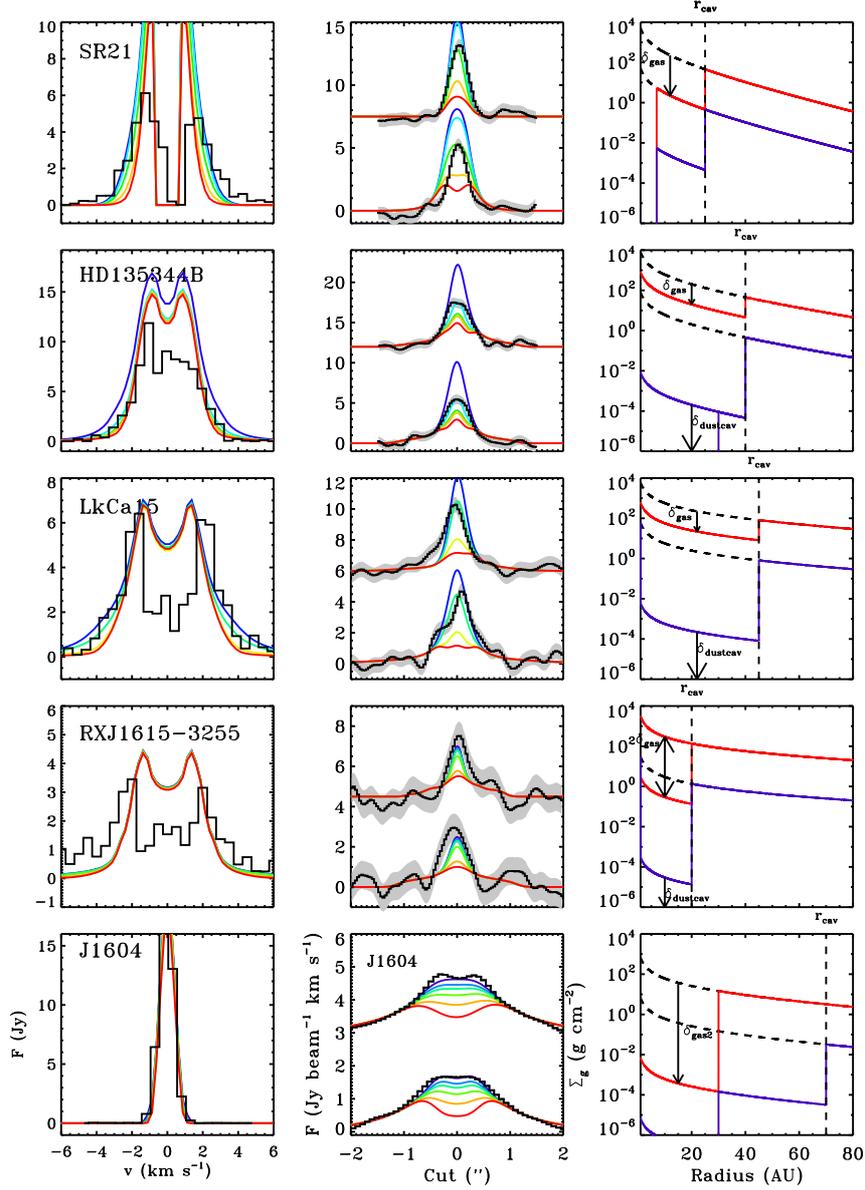}\\
\end{center}
\caption{Modeling results and observations of the $^{12}$CO emission for the best fitting physical model for gas and dust (Figure \ref{fig:gasresults}) comparing $\delta_{\rm gas}$ (gas density drop inside the dust cavity). $\delta_{\rm gas}$ is varied as 10$^{0}$, 10$^{-1}$, ... , 10$^{-4}$ as blue, light-blue, green, yellow and red in the left and middle plots, while the observational data is in black. {\bf Left:} The $^{12}$CO spectrum integrated over the disk area. {\bf Middle:} The intensity cuts through the major (bottom) and minor (top) axis of the $^{12}$CO zero moment map for different $\delta_{\rm gas}$. The noise level is indicated by the gray area. The model images are convolved with the same beam as the ALMA observations. {\bf Right:} The best fitting result for possible values of $\delta_{\rm gas}$ and $\delta_{\rm dustcav}$ in red and blue respectively.}
\label{fig:deltagasresults}
\end{figure*}

\begin{table*}\small
\caption{Results fitting procedure for the gas density profile of each transition disk.}
\label{tbl:fitting}
\begin{tabular}{lllllllllllllll}
\hline
\hline
Target&$r_c$&$\Sigma_c$&$M_{\rm dust}$&$M_{\rm gas}$&$h_c$&$\psi$&$\delta_{\rm dust}$&$r_{\rm cav}$&$r_{\rm sub}$&$r_{\rm gap}$&$f_{\rm ls}$&$r_{\rm out}$&$\delta_{\rm gas}$&$\delta_{\rm dustcav}$\\
&(AU)&(g cm$^{-2}$)&($10^{-3}M_{\odot}$)&($10^{-3}M_{\odot}$)&(rad)&&&(AU)&(AU)&(AU)&&(AU)&&\\
\hline
SR21&15&400&0.12&12&0.07&0.15&$10^{-6}$&25&0.18&1&0.85&100&$10^{-2}$&$10^{-3}$\\  
HD135344B&25&200&0.17&24&0.15&0.05&$10^{-2}$&40&0.18&0.25&0.95&150&$10^{-1}$&$<10^{-4}$\\ 
LkCa15&85&34&0.95&103&0.06&0.04&$10^{-5}$&45&0.08&1&0.98&400&$10^{-1}$&$<10^{-4}$\\ 
RXJ1615-3255&115&60&3.5&470&0.04&0.2&$10^{-6}$&20&0.08&1&0.85&200&$>10^{-4}$&$<10^{-5}$\\ 
SR24S&15&1200&0.35&-&0.12&0.01&$10^{-4}$&25&0.14&1&0.98&-&-&$<10^{-4}$\\ 
J1604-2130&60&12&0.095&20&0.065&0.68&$10^{-1}$&70&0.04&0.06&0.98&400&$10^{-5}$\tablefootmark{a}&$<10^{-6}$\\ 
\hline
\end{tabular}
\tablefoot{
\tablefoottext{a}{This value refers to the drop inside the inner cavity of 30 AU ($\delta_{\rm gas2}$ in the text).}
}
\end{table*}
Our best fitting results for the dust and gas density structure are presented in Figures \ref{fig:dustresults} to \ref{fig:deltagasresults} and Table \ref{tbl:fitting}. In Figure \ref{fig:dustresults} the output of the physical model is compared to the SED, the 690 GHz continuum visibility curve and the normalized major/minor axis cuts of the 690 GHz continuum image (left panel of Figure \ref{fig:intensity}). Data points of the SED have been dereddened using the extinction law by \citet{WeingartnerDraine2001} with $R_V$=5.5. The SED constrains the vertical structure of the disk, whereas the visibility curve determines the cavity size and extent through the location of the null. The image cuts show the direct comparison of the image with model. Within each figure, the $\delta_{\rm dustcav}$ value (drop of dust density inside the cavity) is varied with values between 10$^{-2}$ and 10$^{-6}$, to constrain the minimum drop in dust density inside the cavity. In the right most panel the density structure of the model is presented. 

Reasonable fits are found for the continuum visibilities and SED within the constraints of our physical model for all targets. The inferred disk masses are comparable to previous findings within a factor of a few, e.g. \citet{Andrews2011}. 
The models with $\delta_{\rm dustcav}$=10$^{-2}$ overproduce the emission inside the cavity in all cases, as seen both in the mid infrared part of the SED and in the dust continuum image. Therefore, $\delta_{\rm dustcav}$ is constrained to be at most a factor of 10$^{-4}$, and even as low as 10$^{-6}$ for J1604-2130. In the modeling of the gas, we set $\delta_{\rm dustcav}$ to zero, except for SR21 where $\delta_{\rm dustcav}$ is constrained by the visibilities. For the gas modeling of HD135344B and J1604-2130, we include some additional dust between $r_{\rm cav}$ and 30 AU as shown in the dust density profile in Figure \ref{fig:deltagasresults}. 

For the analysis of the gas density, we compare the result of the adopted disk model to the $^{12}$CO spectrum, the intensity cuts through the zero moment map and the $^{12}$CO visibility curve in Figure \ref{fig:gasresults}. SR24S is not included in this part of the analysis because the CO emission is too heavily confused by large scale emission (Fig. \ref{fig:channelmap}). The models are calculated assuming $\delta_{\rm gas}$=1, i.e. no drop in the gas surface density inside the dust cavity, except for J1604-2130, which shows a resolved drop in the CO emission inside 30 AU, which is adopted already. Because the CO visibilities do not show a clear null and the disks are not fully symmetric in the gas, the visibility curve was not used to fit our data exactly: it is merely included here to show that the emission from the moment map is well retrieved from the visibilities. 

The models reproduce the CO emission well in the outer part of the disk ($r>r_{\rm cav}$). There was no need to change the gas to dust ratio of 100 in any of the cases in the outer disk. This conclusion needs to be confirmed by optically thin CO isotopologue observations. The images and visibilities show that the outer extent of the dust and CO emission can indeed be fit with the same physical model, although better $S/N$ observations are required to confirm this. 

For SR21, HD135344B and LkCa15, the initial model overproduces the central part of the intensity cut by a factor 1.5--2, suggesting a decrease of gas surface density inside the dust cavity. In Figure \ref{fig:deltagasresults} we present the same physical model with $\delta_{\rm gas}$ varying from 10$^{0}$ to 10$^{-4}$ in steps of 10 to compare with the observations. The emission drops by less than a factor of 2 in each of these steps due to the optical depth of the $^{12}$CO line, but within the noise  level of the moment map this is sufficient to constrain $\delta_{\rm gas}$ to within an order of magnitude due to the different gas temperatures at the $\tau=1$ surface in the disk. This result is different from Oph IRS 48, where the $^{12}$CO line wings were found to be optically thin \citep{Bruderer2014}. Only the emission in J1604-2130 inside 30 AU radius becomes optically thin. The range of $\delta_{\rm gas}$ is indicated in the right panel of Figure \ref{fig:deltagasresults} and in Table \ref{tbl:fitting}. 

In the following section we discuss the properties of each disk in more detail.

\subsection{Results of individual targets}

\paragraph{SR21\\}
It is possible to fit the SED and amplitudes of the continuum visibilities simultaneously at least up to baselines of 600 k$\lambda$. The small inconsistency at longer baselines is possibly related to the dust asymmetry \citep{Perez2014} which is not taken into account. In order to fit the SED, we use for this disk an alternative dust opacity table for the small grains, without small silicates, because the normal dust table results in a strong silicate feature that is not seen in the data. We find a cavity size of at most 25 AU, which is significantly smaller than the value of 36 AU previously found in the SMA data at 345 GHz \citep{Brown2009, Andrews2011}. It remains unclear whether this is due to differences in the modeling approach or a physical effect, since the 345 GHz emission traces somewhat larger dust grains (see also 5.2). Similar to \citet{Andrews2011} we find that this disk has a remarkably small radius: $r_c$ is only 15 AU, resulting in no detectable emission outside of 75 AU radius. 

Both hot CO line emission \citep{Pontoppidan2008} and mid infrared interferometry (Benisty, private comm.) suggest a hot ring at 7 AU and no material inside. Also scattered light shows that there is small dust down to $<$14 AU  \citep{Follette2013}. Therefore, we assume that there is no gas between the star and 7 AU for SR21 (but still an inner dust disk due to the NIR excess), and dust with only a small grain population between 7 AU and $r_{\rm cav}$. The dust between 7 AU and the dust hole cavity of 25 AU is partly constrained by the visibility curve: without the additional dust the second null is much further out, although this may also be due to the asymmetry. In constraining $\delta_{\rm dustcav}$, this drop is assumed to start from 7 AU rather than from $r_{\rm gap}$ (taken as 1 AU).

The central $^{12}$CO emission in the intensity cut is overproduced by about a factor of 2 in the initial model, especially inside the cavity region (Fig. \ref{fig:gasresults}). Only a deep drop in $\delta_{\rm gas}$ of 2 orders of magnitude can match the peak of the zero moment map, although the intensity cut is somewhat wider than the data. A minor dip of the 7 AU gap is seen in the intensity cut of the model integrated $^{12}$CO emission: the gap is marginally resolved, but the dip is within the error bars of the observations so it cannot be confirmed with these data. 

\paragraph{HD135344B\\}
The continuum visibility is reasonably well fitted with a model with a 40 AU cavity, slightly smaller than the 45 AU cavity found in previous work with 345 GHz SMA data \citep{Brown2009,Andrews2011,Carmona2014}. An increase of the large grain fraction and large grain scale height ($\chi=0.8$) is needed to fit the peak of the far infrared part of the SED. 

We find a smaller critical radius of only 25 AU and thus smaller extent of the disk compared to \citet{Andrews2011} (who found $r_c$=55 AU), but the ALMA data quality is much better than the SMA data. Another main difference in the modeling procedure is the different vertical structure of this disk in our parametrization (no puffed up inner rims), although this is not expected to change the emission in the outer disk. 
Scattered light images show that there is small dust inside the millimeter dust cavity down to 30 AU for HD135344B (millimeter dust cavity is $\sim$46 AU) \citep{Garufi2013}. Therefore we assume a finite drop in dust density between $r_{\rm cav}$ and 30 AU. The scattered light images, SED and visibilities do not constrain the amount of dust: we choose a drop of 10$^{-3}$. \citet{Garufi2013} and \citet{Muto2012} show signatures of spiral arms in the scattered light images. The ALMA data have insufficient resolution to detect these arms if they were visible at millimeter wavelengths.

\begin{figure}[!ht]
\includegraphics[scale=0.4]{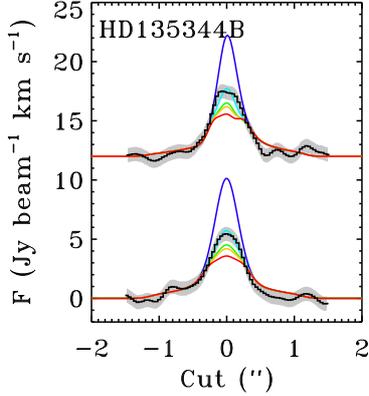}
\caption{Modeling results and observations of the $^{12}$CO emission for the best fitting physical model for HD135344B for a 30 AU rather than 40 AU cavity for gas and dust (Figure \ref{fig:gasresults}) for different values of $\delta_{\rm gas}$ (gas density drop inside the dust cavity). $\delta_{\rm gas}$ is varied as 10$^{0}$, 10$^{-1}$, ... , 10$^{-4}$ as blue, light-blue, green, yellow and red. The observations are plotted in black. The panel shows the intensity cuts through the major (bottom) and minor (top) axis of the $^{12}$CO zero moment map.}
\label{fig:HD13COgap}
\end{figure}

The initial model for this disk overproduces the CO emission inside the cavity. A $\delta_{\rm gas}$ of 10$^{-1}-10^{-2}$ is sufficient to fit the peak in the moment map. A decrease in gas density inside the cavity was previously inferred from SMA CO line observations \citep{Lyo2011}. \citet{Carmona2014} require a surface density that increases with radius inside the cavity to fit the rovibrational CO lines in this disk. Fitting these lines is beyond the scope of this paper, but the presence of a drop in surface density may be similar to this increasing density profile, as discussed in Section 5.1. 
Motivated by the dust structure we also try a gas model with a cavity size of 30 AU rather than 40 AU. It turns out that there is no significant difference in the outcome for 30 AU or 40 AU gas cavity radius. As Figure \ref{fig:HD13COgap} shows, $\delta_{\rm gas}=10^{-1}-10^{-2}$ also fits best for a 30 AU gap.

\paragraph{LkCa15\\}
LkCa15 is a well studied transition disk with a large cavity of 50 AU \citep{Isella2012}. We model LkCa15 with a rather flat but massive disk structure. Our derived disk dust mass is 3 times higher than typically found in the literature, and the 690 GHz continuum is indeed optically thick, which explains the apparent asymmetry in the continuum image: the north west side of the disk is brighter than the south east due to the geometry. 

Just like HD135344B, the initial model for this disk overproduces the CO emission inside the cavity, indicating a $\delta_{\rm gas}$ of at least 10$^{-1}$. 

\begin{figure}[!ht]
\includegraphics[scale=0.4]{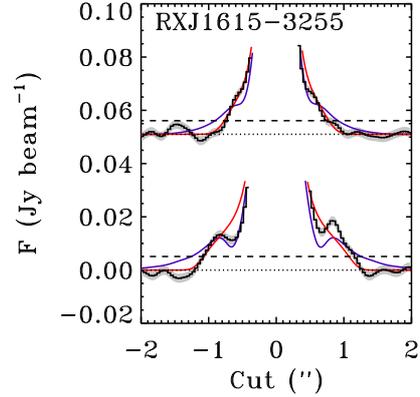}
\caption{Modeling results and observations of the 690 GHz continuum of RXJ1615-3255. The panel shows the intensity cuts through the major (bottom) and minor (top) axis of the continuum image, zoomed in on the lowest flux levels ($<20\sigma$). The observations are plotted in black with the 1$\sigma$ errors in gray. In red and blue the models with and without a dust gap between 110 and 130 AU are presented. The dotted and dashed lines indicates the zero flux and 3$\sigma$ limits, respectively.}
\label{fig:zoomrxj1615}
\end{figure}

\paragraph{RXJ1615-3255\\}
RXJ1615-3255 does not reveal a cavity in the continuum image, and also the visibilities show only a hint of a null at the longer baselines ($>$500 k$\lambda$). Our best fitting model has a 20 AU cavity, somewhat smaller than the best fit in \citet{Andrews2011} of 30 AU. A model without cavity can not fit the SED which shows a clear indication of a deficit of dust close to the star. This indicates that the Band 9 continuum might be optically thick. An interesting aspect of this target is the outer part of the disk: a narrow radial dip is seen in the image at 0.6'' from the center (corresponding to 110 AU at the distance of this disk). Figure \ref{fig:zoomrxj1615} shows a zoomed-in image of the intensity cut from Figure \ref{fig:dustresults}. This hints at the presence of a dust gap in the outer part of the disk between 110 and 130 AU, or an additional outer ring of dust, as seen in recent ALMA data of HD100546 \citep{Walsh2014}.

The model with $\delta_{\rm gas}$=1 does not reproduce the high velocity emission seen in the CO spectrum, but the $S/N$ is low so the significance of this remains debatable. The intensity cuts show that the gas density inside the dust hole can be up to 4 orders of magnitude lower before the CO becomes optically thin. 

\paragraph{SR24S\\}
SR24S is an inclined disk with a just resolved cavity of about 30 AU. The long wavelength part of the Spitzer IRS spectrum and the MIPS-24 flux are known to be confused by its nearby companion SR24N (see e.g. \citealt[][]{Andrews2011}), thus these points are only used as upper limits in our fit. The visibility curve becomes inconsistent with our model for baselines $>$350 k$\lambda$, hinting at substructure inside the cavity. As we do not have any additional information on dust inside the cavity of this disk we do not explore this further. 

\paragraph{J1604\\}
J1604-2130 is a well-studied transition disk at 345 GHz, both with SMA \citep{Mathews2012} and ALMA  \citep{Zhang2014}. Both authors provide physical models for both the dust and the gas. \citet{Zhang2014} uses the same ALMA dataset discussed here, but does not use a full physical-chemical model for the gas analysis. Also, in both studies the near infrared excess is not taken into account. The issue is the apparent variability: whereas the Spitzer IRAC photometry (taken in 2006 and used by Mathews) shows no near infrared excess, the Spitzer IRS and WISE photometry (taken in 2007 and 2012, respectively) do reveal the excess. Zhang et al. note this discrepancy and state that the near infrared excess can be fit with an optically thick inner disk ring, but do not include this in their model. Since the inner disk has a strong influence on the CO emission inside the cavity (and on the shadowing of the dust cavity wall), we choose to include the more recent photometry with near infrared excess in our SED and fit the models accordingly. 

\citet{Mathews2012} introduce a two drop dust density model to fit their SMA data which is also adopted by \citet{Zhang2014} in the analysis of the ALMA data: one drop at 75 and one at 20 AU. We find that such a two-step decrease is indeed required to fit the continuum visibilities and adopt the same structure, although we set the inner drop radius at 30 AU to be consistent with the gas cavity. On the other hand, we require a much larger critical radius of 60 AU rather than 10-20 AU as used by Mathews and Zhang. This is likely the result of shadowing due to the presence of an inner disk. Our disk dust mass of 0.1 $M_{\rm Jup}$ is comparable with their estimates. 

\begin{figure}[!ht]
\includegraphics[scale=0.3]{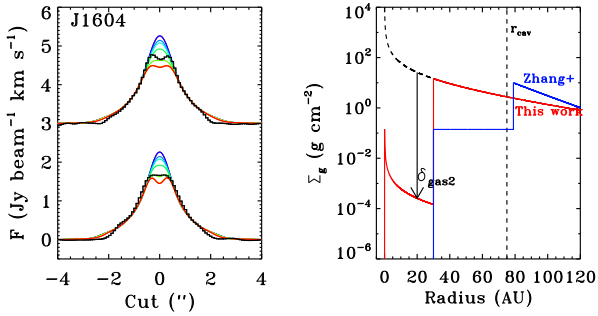}
\caption{{\bf Left:} Modeling results and observations of the $^{12}$CO emission for the best fitting physical gas model for J1604-2130 for the inner 30 AU cavity comparing $\delta_{\rm gas2}$ (gas density drop inside the 30 AU dust cavity). $\delta_{\rm gas2}$ is varied as 10$^{0}$, 10$^{-1}$, ... , 10$^{-6}$ as blue, light-blue, cyan, green, yellow,orange and red ($\delta_{\rm gas}$ is taken as 1). The observations are plotted in black. The panel shows the intensity cuts through the major (bottom) and minor (top) axis of the $^{12}$CO zero moment map. {\bf Right:} Comparison between the density profile derived in this study and the profile used in \citet{Zhang2014}.}
\label{fig:J1604COgap}
\end{figure}

In the gas a clear signature for a drop at 30 AU is seen in the CO moment map, which was modeled by \citet{Zhang2014} in a parametrized gas model of a power-law temperature profile with an empty gap. Their result is consistent with that of our full physical-chemical model and the derived surface density profile is similar (right panel of Figure \ref{fig:J1604COgap}). We can constrain the inner density drop to $\delta_{\rm gas2}=10^{-5}$ (see Figure \ref{fig:J1604COgap}), although this depends on the inner cavity size: a smaller cavity size would require an even deeper drop, a larger size a less deep drop. On the other hand, $\delta_{\rm gas}$ (the drop inside the dust cavity radius) is 1, there is no visible drop in density between 30 and 70 AU.

\section{Discussion}
\subsection{Implications of gas density drop}
The key parameter in our analysis is $\delta_{\rm gas}$, the drop in gas density inside the dust hole, in comparison with $\delta_{\rm dustcav}$. For each target, the minimum drop in dust density $\delta_{\rm dustcav}$ is significantly lower than the possible values of $\delta_{\rm gas}$. Figure \ref{fig:deltagasresults} and Table \ref{tbl:fitting} show the range of possible values for these parameters for each target. 

The significantly larger drop in dust density compared with the gas is consistent with the dust trapping scenario by a companion, with gas present inside the millimeter-sized dust cavity \citep{Pinilla2012b}. For the sources studied here, photoevaporation by itself is unlikely based on just the large cavity sizes and high accretion rates \citep{Owen2011,Rosotti2013} and our results further strengthen this hypothesis. 
Especially SR21, HD135344B and LkCa15 require a minimum drop in the gas surface density of a factor 10-100 within the gap and are thus promising candidates for disks with embedded planets. The drop is deep enough to rule out the dead zone scenario, which has a drop that is much less than an order of magnitude \citep{Lyra2015}. On the other hand, one has to be careful with overinterpretation at this point: the CO emission inside the gap is sensitive to the amount of small dust inside the cavity due to its shielding effect, which can also explain the apparent drop in CO emission. We have explored this within the constraints of our physical model, but any addition of dust or PAHs inside the cavity that is sufficient to lower the CO emission at the same time increases the mid infared continuum emission in the SED to levels that are inconsistent with the data.  The combination of SED modeling and spatially resolved $^{12}$CO emission thus provide unique constraints on the temperature and small dust grain composition. On the other hand, the vertical structure also influences the temperature structure and thus the $^{12}$CO emission and mid infrared continuum emission, and this degeneracy remains uncertain.

The depth of the gas density drops at the cavity radius is still modest, only a factor $\delta_{\rm gas}\sim 10^{-1}-10^{-2}$. Comparison with the planet-disk interaction models of Jupiter mass planets in Figure 1 in \citet{Pinilla2012b} indicates that any embedded planets responsible for this drop are unlikely to be more massive than 1-2 Jupiter masses, and even then only with high viscosity ($\alpha\sim10^{-2}$). A high viscosity is inconsistent with the presence of long-lived vortices \citep[e.g.][]{Ataiee2013}. \citet{Perez2014} interpret the asymmetries of SR21 and HD135344B as possible vortices, by comparing their analytic solution for the asymmetric dust structures with the vortex prescription of \citet{LyraLin2013}. The lack of a deep gap in the CO argues against this interpretation. \citet{Pinilla2015} also conclude that for SR 21 the vortex scenario is unlikely by analyzing the observed dust distribution at different wavelengths with hydrodynamic and dust evolution models. However, a smaller gas cavity size with a deeper density drop would also fit our observations and be consistent with low viscosities. Further high spatial resolution observations of CO isotopologues are needed to disentangle cavity size and gas density drop.

The planet-disk interaction models actually predict rounded off gradients rather than steep vertical drops such as parametrized in our model. A rounded-off dust wall was indeed inferred from mid infrared interferometry data of HD~100546 \citep{Mulders2013} by fitting the visibility curve, but the S/N of our data is insufficient to test this. One important conclusion of \citet{Pinilla2012b} is that the gas cavity radius (location of the planet) is expected to be up to a factor of 2 smaller than the dust cavity radius (location of the radial pressure bump where the dust is trapped). This was indeed observed for J1604-2130 (\citealt{Zhang2014}, this study), Oph IRS 48 \citep{Bruderer2014} and HD~142527 \citep{SPerez2014}. As stated above, our estimates of the drop in gas density inside the cavity would likely increase if we would assume a smaller gas cavity size, but the spatial resolution of our data is insufficient to constrain this. For HD135344B we show the outcome for a gas cavity of 30 rather than 40 AU (Fig. \ref{fig:HD13COgap}), but for this radius the $\delta_{\rm gas}$ is still limited to 10$^{-1}-10^{-2}$. For J1604-2130, we find clear evidence for an inner drop at 30 AU, and no additional drop between 30 and 70 AU of 10$^{-1}$, as found for Oph IRS 48 \citep{Bruderer2014}. A structure with a double drop or smaller gas cavity size hints at the presence of more than one companion, or alternatively a shallow increasing slope such as seen in \citet{Pinilla2012b,Pinilla2015}. This possibility can not be excluded for any of the other sources in this study with the available observations. \citet{Carmona2014} suggest a gas surface profile increasing with radius inside the cavity to match the gas and dust density distribution of HD135344B, although this is only constrained by observations for the inner few AU. Modeling the shape of the gap is beyond the scope of this paper. Spatially resolved observations of CO isotopologues will provide better constraints on the density profile. 

\subsection{Cavity size and dust distribution}
Another interesting aspect of our results is that all disks have significantly smaller cavity sizes (by 5-10 AU) in the 690 GHz continuum data than those derived by \citet{Andrews2011} based on the 345 GHz continuum observations taken with the SMA. The cavity size as derived with the ALMA continuum visibilities (location of the null) is constrained to better than $\pm$5 AU. The $S/N$ of the SMA data is much lower, so the cavity size is less precisely constrained but the systematic trend is significant. For comparison, we list the location of the first null for both the 690 GHz and 345 GHz continuum of our adopted model in Table \ref{tbl:nulls} and compare those with the null in \citet{Andrews2011}. 

\begin{table}\small
\caption{Location of the null in the visibilities of models and SMA observations at various frequencies \citep{Andrews2011}.}
\label{tbl:nulls}
\begin{tabular}{llll}
\hline
\hline
Target&Model-690&Model-345&Observed-345\\
&null (k$\lambda$)&null (k$\lambda$)&null (k$\lambda$)\\
\hline
SR21&252&247&200-220 \\
HD135344B&196&204&160--180 \\ 
LkCa15&171&155&140-160 \\
RXJ1615-3255&484&349&250--300 \\
SR24S&255&260&240--260 \\
\hline
\end{tabular}
\end{table} 

Note that the null as derived from the models can shift slightly for the two frequencies, due to difference in dust opacity at the observed wavelengths. The null in the observed 345 GHz continuum is significantly smaller for SR21, HD135344B and RXJ1615-3255 than that at 690 GHz, indicating a larger cavity size at this longer wavelength. This hints at a radial variation of the dust size distribution, as the continuum emission is sensitive to larger grains at longer wavelengths. A more concentrated dust ring (i.e. a narrower ring and thus a larger dust cavity size) is consistent with the dust trapping scenario, because larger dust grains are further decoupled from the gas and thus more affected by radial drift and trapping \citep{Brauer2008}. 
Higher S/N ALMA Band 7 observations are required to confirm this. 

\section{Conclusions}
In this work, we have analyzed high spatial resolution ALMA submillimeter observations of $^{12}$CO line emission from 6 transitional disks using a physical-chemical model. By comparing the SED, the 690 GHz continuum visibilities and the $^{12}$CO emission simultaneously, we derive a physical model of the gas and dust. With the model we can set constraints on the surface density profile of the dust and gas and specifically, the amount of dust and gas in the cavity.
\begin{enumerate}
\item All disks show clear evidence for gas inside the dust cavity.
\item The gas and dust observations can be fit with a surface density profile with a steep density drop at the cavity radius, taking a gas-to-dust ratio of 100 in the outer disk.
\item The combination of SED and spatially resolved $^{12}$CO intensity fitting sets constraints on the vertical and thus temperature structure of each disk.
\item All disks except SR24S have a potential drop of 1 or 2 orders of magnitude in the gas surface density inside the mm-sized dust cavity, while the minimal drop in dust surface density inside the cavity is at least 3 orders of magnitude. 
\item J1604-2130 has a deep resolved gas cavity that is smaller than the dust cavity. For the other disks it is possible that the gas cavity radius is smaller than the dust cavity radius, in which case the density drop will be deeper, but this can not be constrained with the available data.
\item The derived density profiles suggest the clearing of the cavity by one or more companions, trapping the millimeter dust at the edge of the cavity.
\item Our model for J1604-2130 is mostly consistent with the proposed physical structure by \citet{Zhang2014}, derived using a parametrized temperature model.
\item The continuum of RXJ1615-3255 shows an additional dust ring around 120 AU.
\item The derived cavity sizes of the millimeter dust (and location of the null in the visibilities) are consistently smaller for the 690 GHz/0.44 mm continuum than for the 345 GHz/0.88 mm continuum. This is consistent with the dust trapping scenario, because trapping is more efficient for larger dust grains probed at longer wavelengths.
\end{enumerate}

The derived physical model can be used as a start for the analysis of CO isotopologue observations, putting better constraints on the gas density inside and outside the cavity. We have recently obtained CO isotopologue ALMA observations for two disks of the sample in this study (van der Marel et al.\ in prep.).

  \begin{acknowledgements}
  The authors would like to thank P. Pinilla, A. Juhasz and C. Walsh for useful discussions and the referee for his constructive comments.
  N.M. is supported by the Netherlands Research School for Astronomy
  (NOVA), S.B. acknowledges a stipend by the Max Planck
  Society. Astrochemistry in Leiden is supported by the Netherlands
  Research School for Astronomy (NOVA), by a Royal Netherlands Academy
  of Arts and Sciences (KNAW) professor prize, and by the European
  Union A-ERC grant 291141 CHEMPLAN. This paper makes use of the
  following ALMA data: ADS/JAO.ALMA/2011.0.00724.S, 2011.0.00526.S and 2011.0.00733.S. ALMA is  a partnership of ESO (representing its member states), NSF (USA) and
  NINS (Japan), together with NRC (Canada) and NSC and ASIAA (Taiwan),
  in cooperation with the Republic of Chile. The Joint ALMA
  Observatory is operated by ESO, AUI/NRAO and NAOJ.  
  \end{acknowledgements}

\bibliographystyle{aa}

\appendix
\section{Channel maps}
\begin{table*}[!ht]
\caption{Velocity range within the cavity radius}
\label{tbl:velocitycav}
\begin{tabular}{ll}
\hline
Target&Velocity (km s$^{-1}$)\\
\hline
SR21&$<$1.25 and $>$4.0\\
HD135344B&$<$5.0 and $>$9.0\\
LkCa15&$<$2.2 and $>$10\\
SR24S&$<$1.25 and $>$8.25\\
RXJ1615-3255&$<$-0.3 and $>$9.5\\
J1604-2130&$<$4.1 and $>$5.3\\
\hline
\end{tabular}
\end{table*}
\begin{figure*}
\subfigure{\includegraphics[scale=0.9]{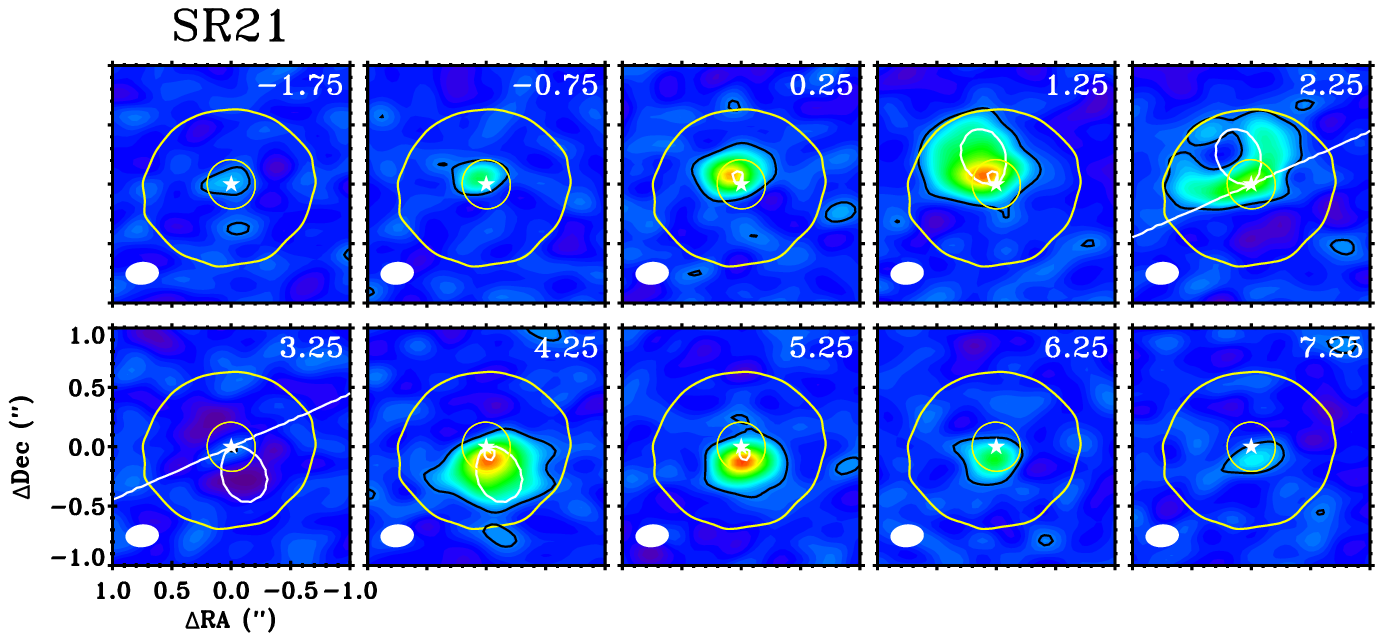}}
\subfigure{\includegraphics[scale=0.4,trim=100 0 0 0]{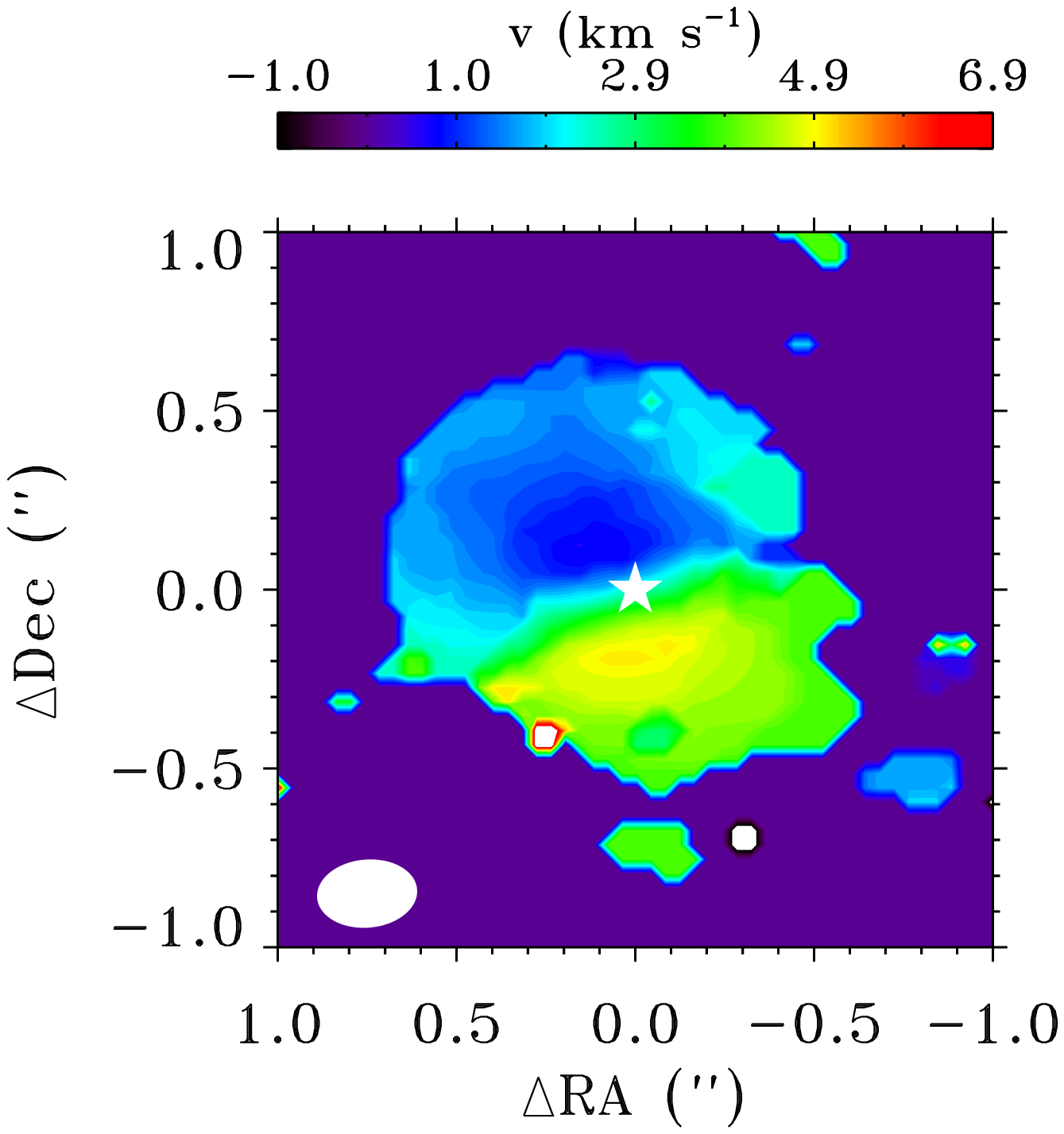}}\\
\subfigure{\includegraphics[scale=0.9]{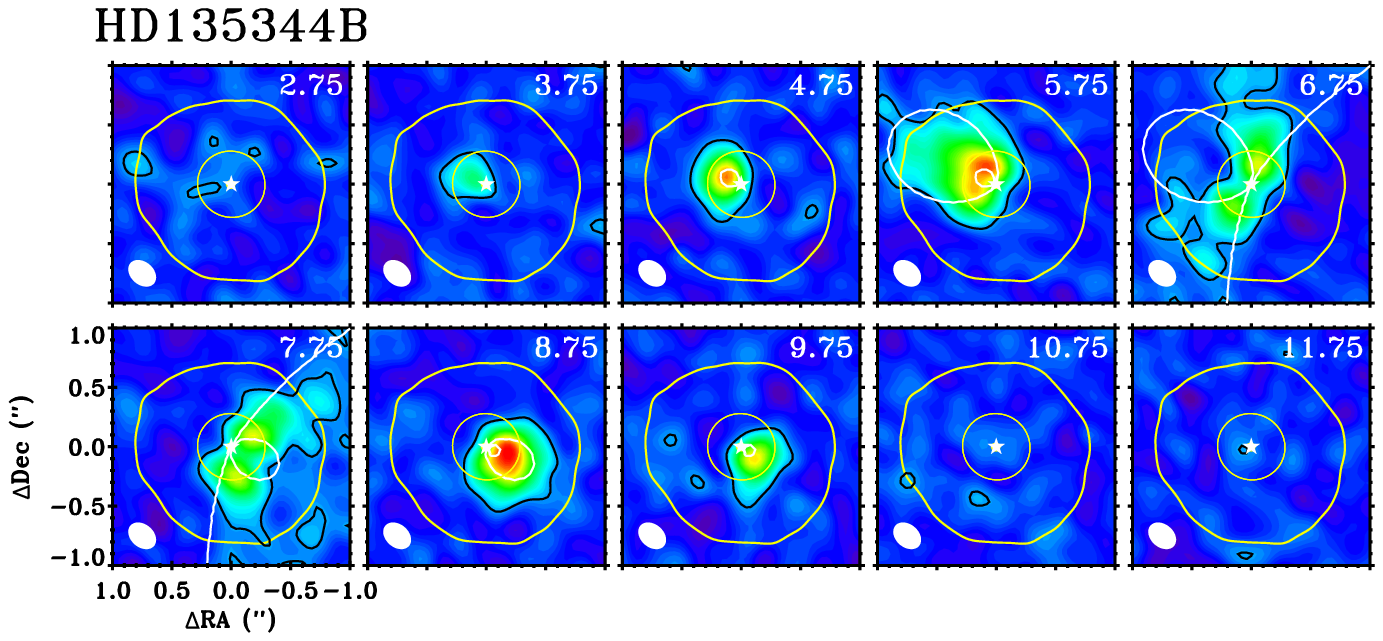}}
\subfigure{\includegraphics[scale=0.4,trim=100 0 0 0]{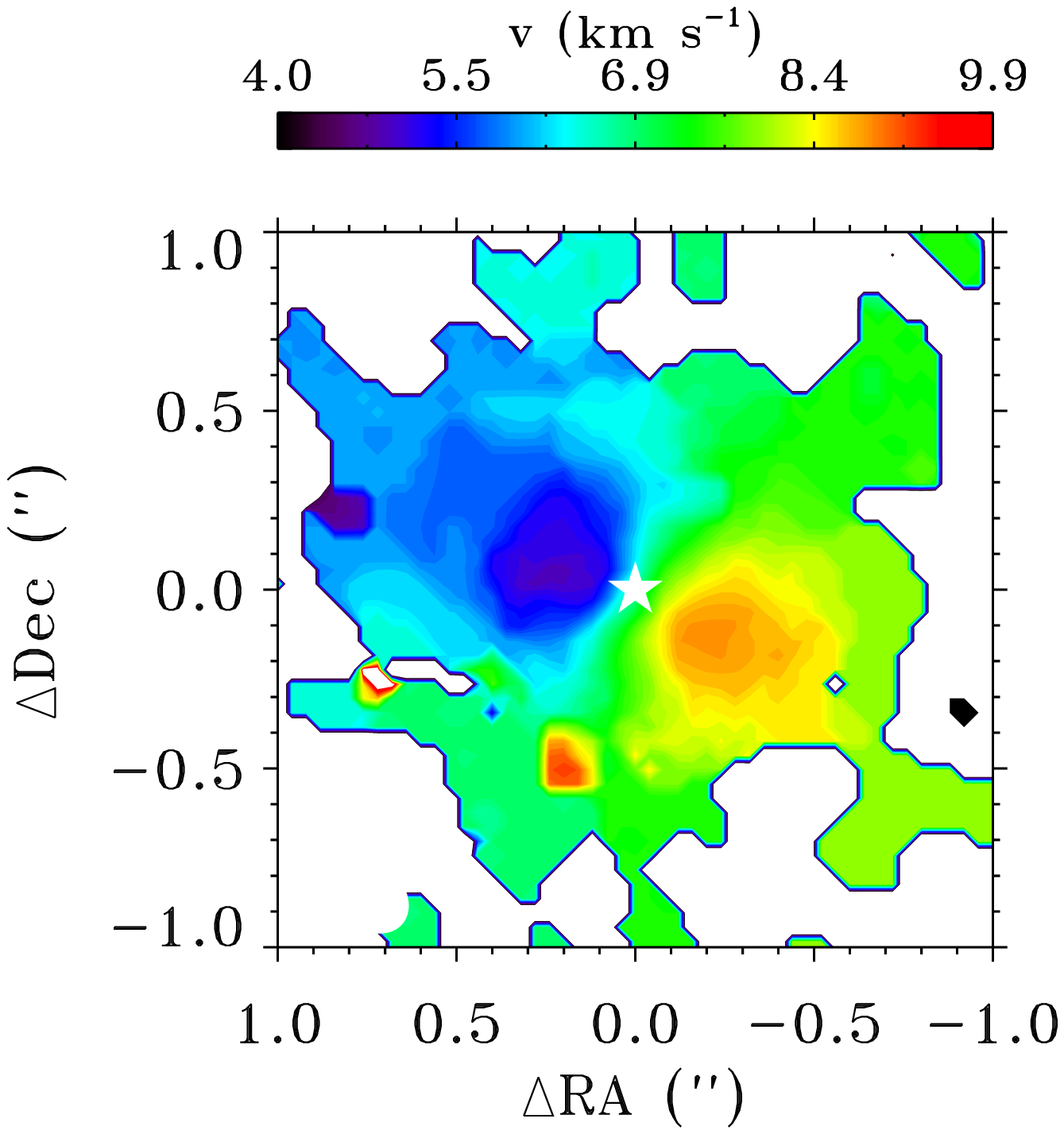}}\\
\subfigure{\includegraphics[scale=0.9]{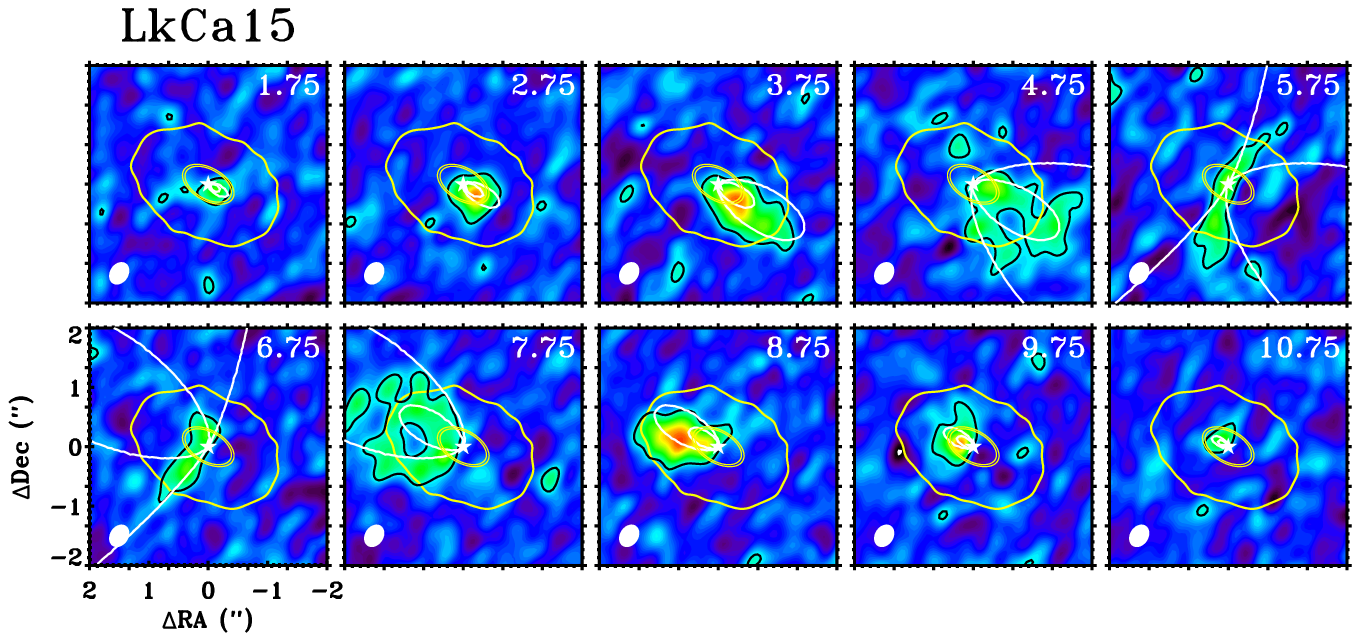}}
\subfigure{\includegraphics[scale=0.4,trim=100 0 0 0]{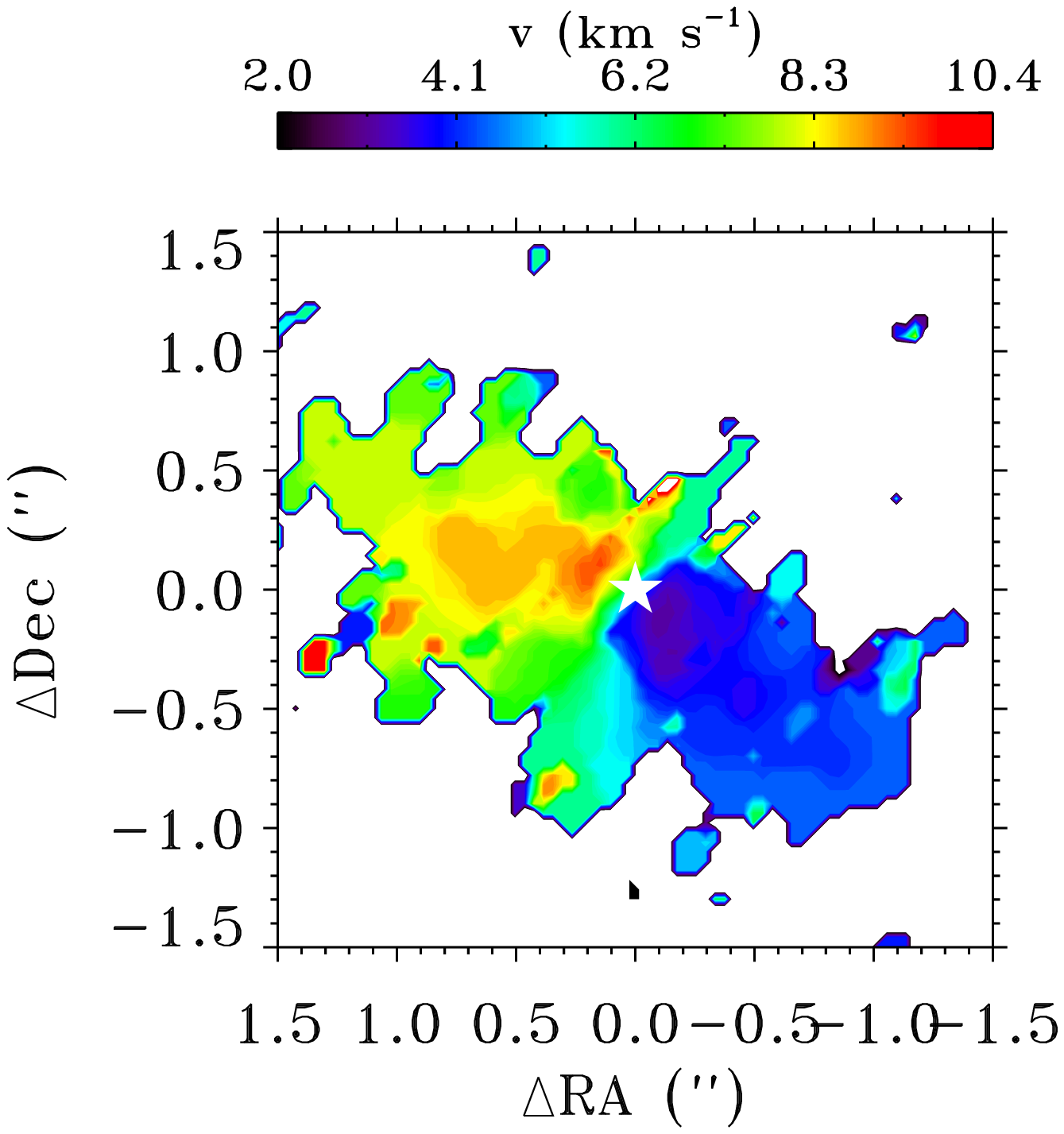}}\\
\captcont{$^{12}$CO channel maps for each observed target. Overlaid in white contours are the Keplerian velocity profiles for the derived inclination and given stellar mass. The yellow ellipse indicates the dust hole radius and the yellow contours the 5$\sigma$ outer radius of the dust continuum. }
\label{fig:channelmap}
\end{figure*}

\begin{figure*}
\subfigure{\includegraphics[scale=0.9]{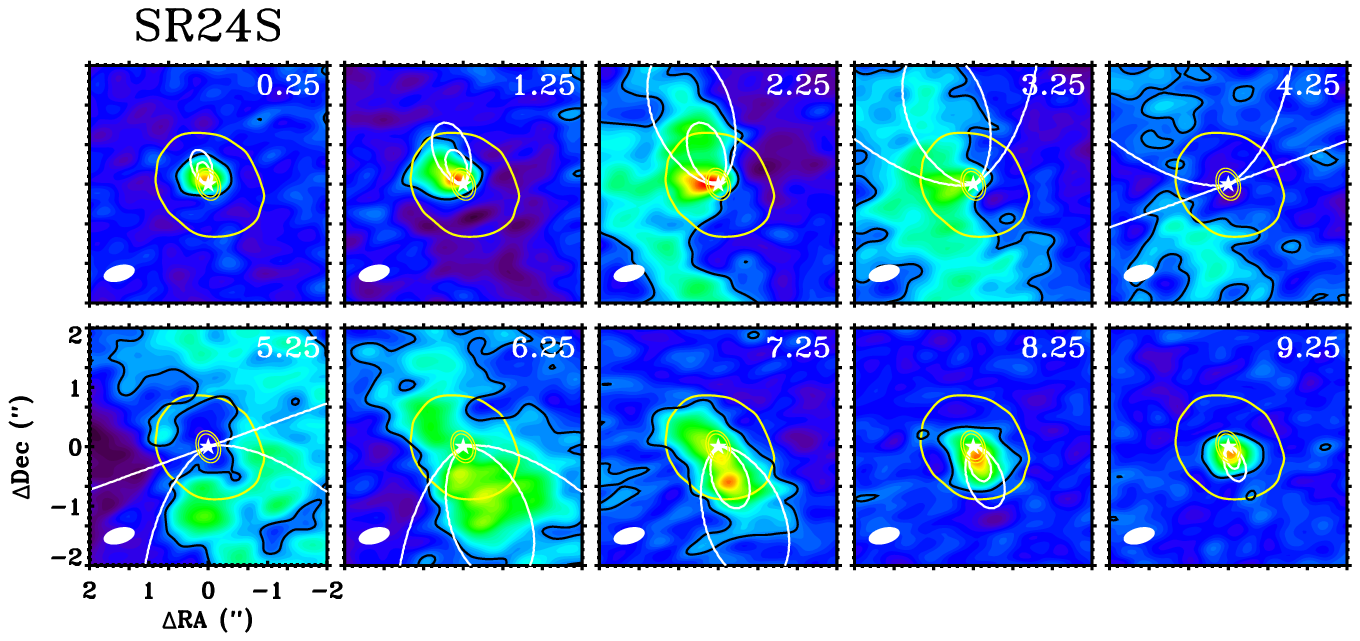}}
\subfigure{\includegraphics[scale=0.4,trim=100 0 0 0]{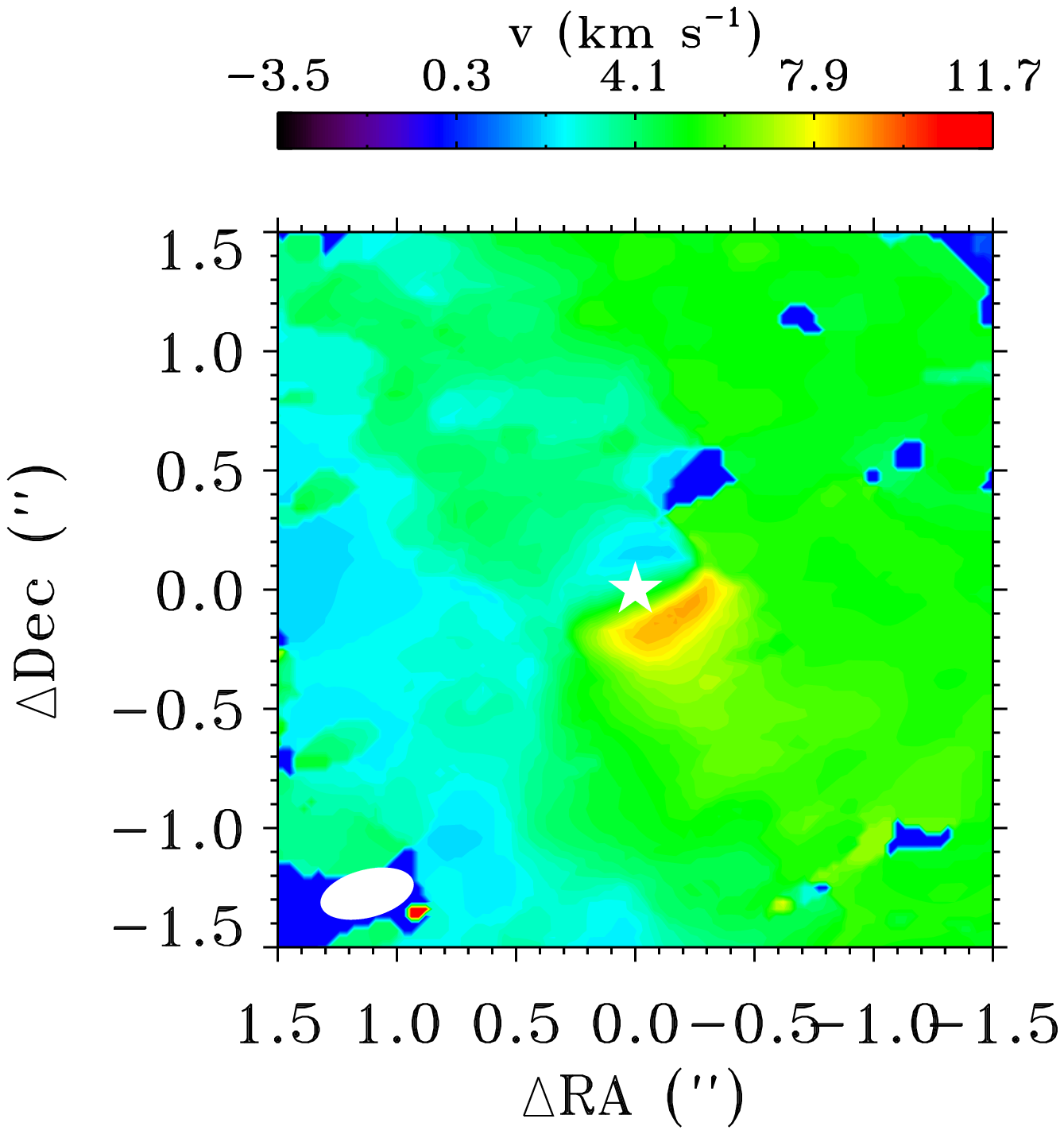}}\\
\subfigure{\includegraphics[scale=0.9]{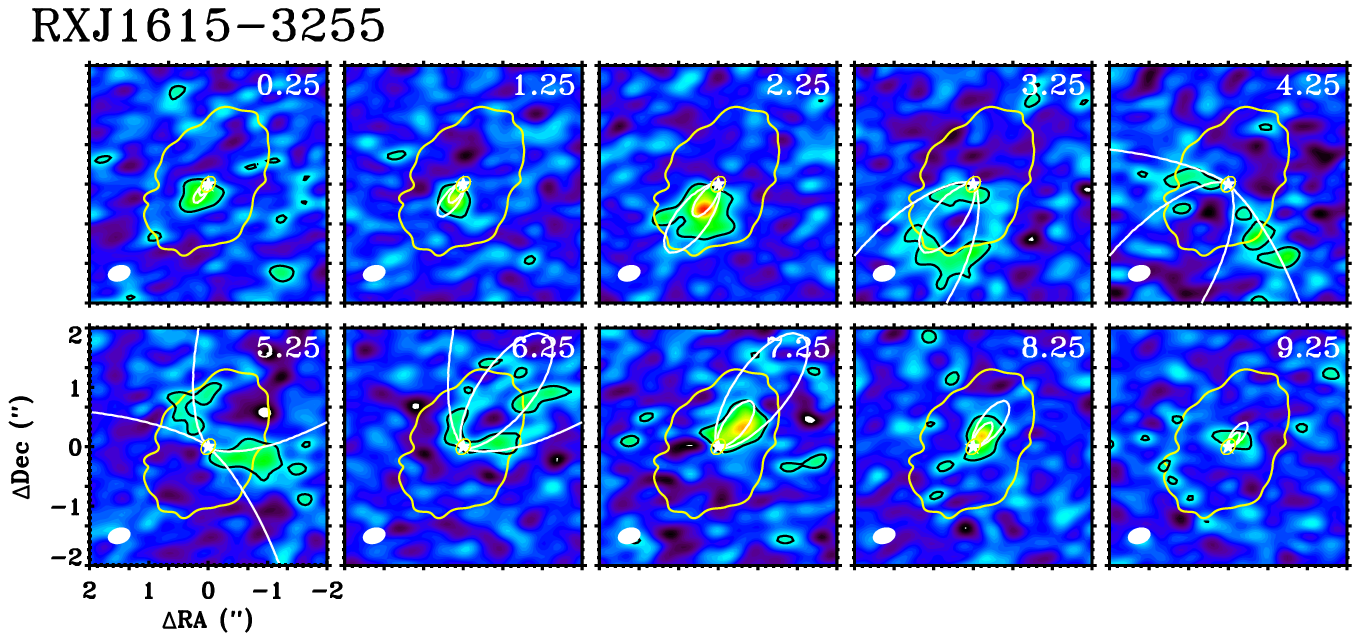}}
\subfigure{\includegraphics[scale=0.4,trim=100 0 0 0]{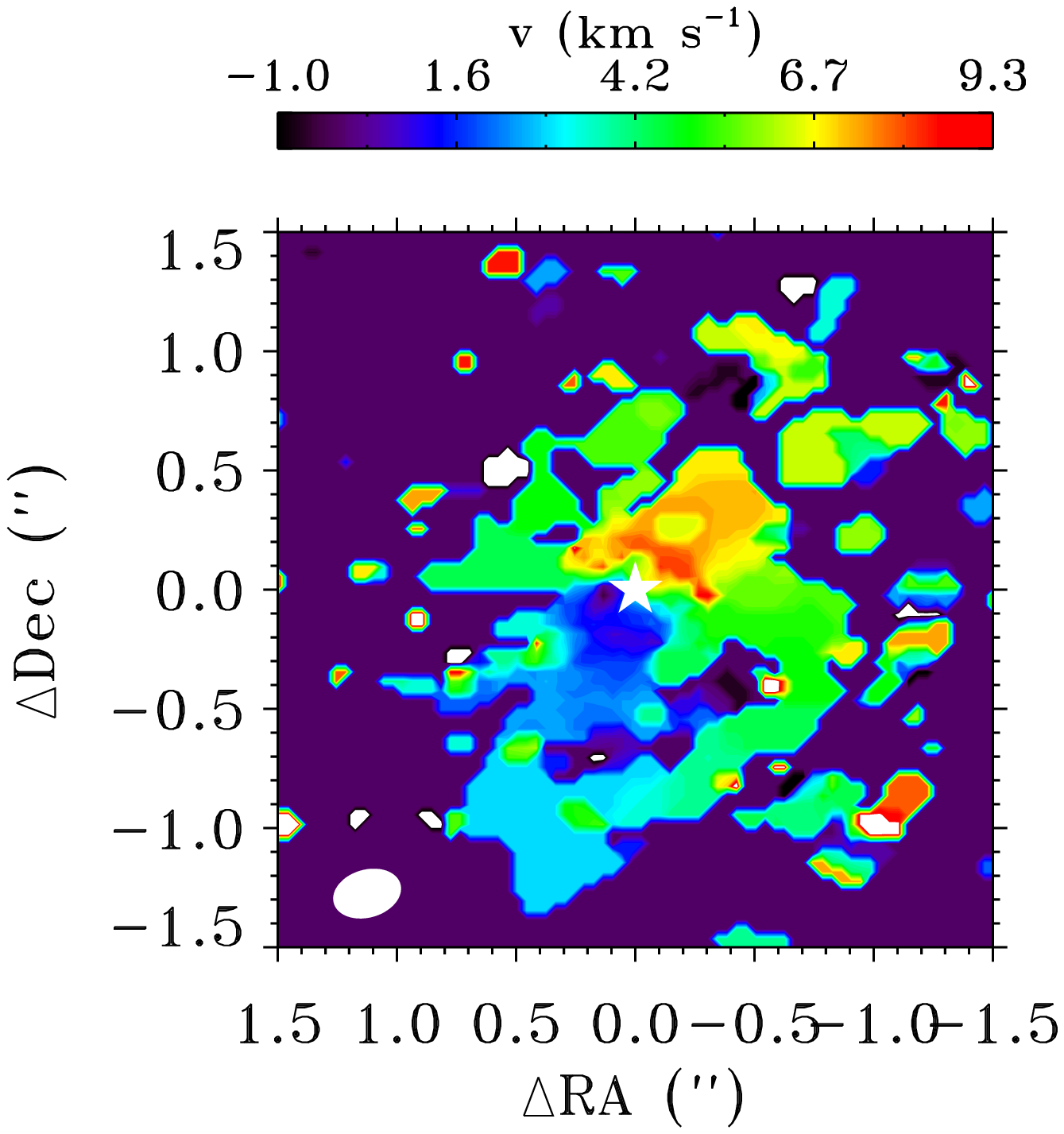}}\\
\subfigure{\includegraphics[scale=0.9]{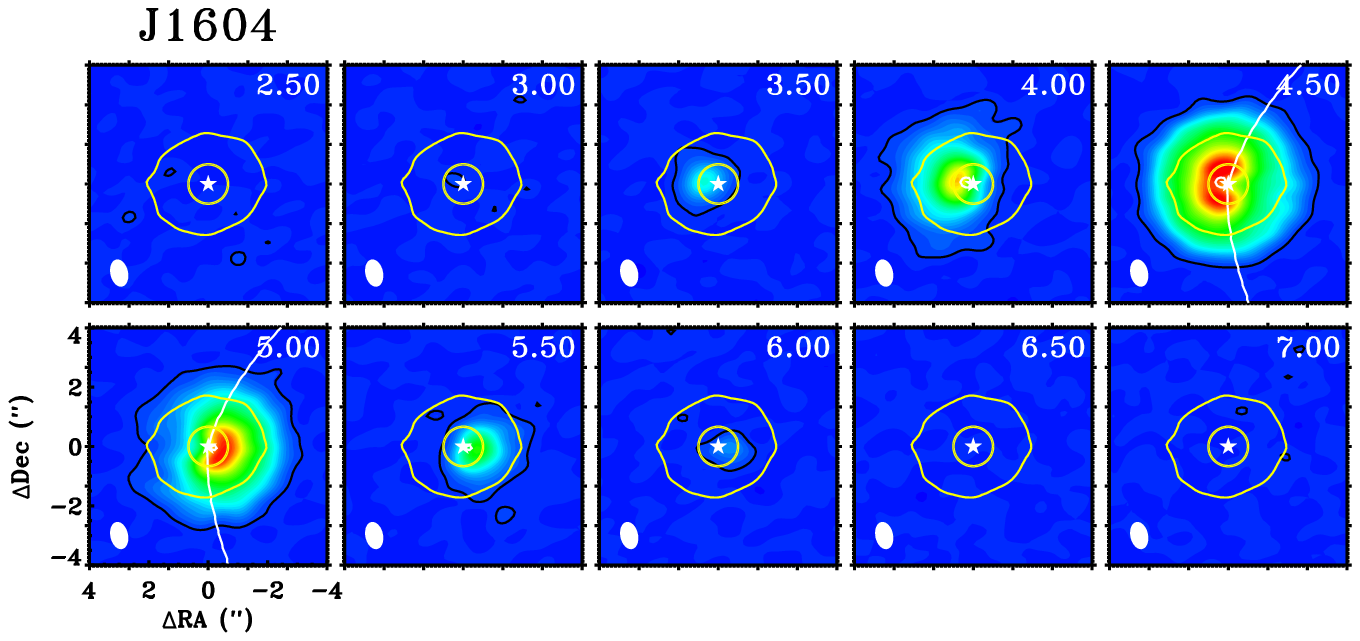}}
\subfigure{\includegraphics[scale=0.4,trim=100 0 0 0]{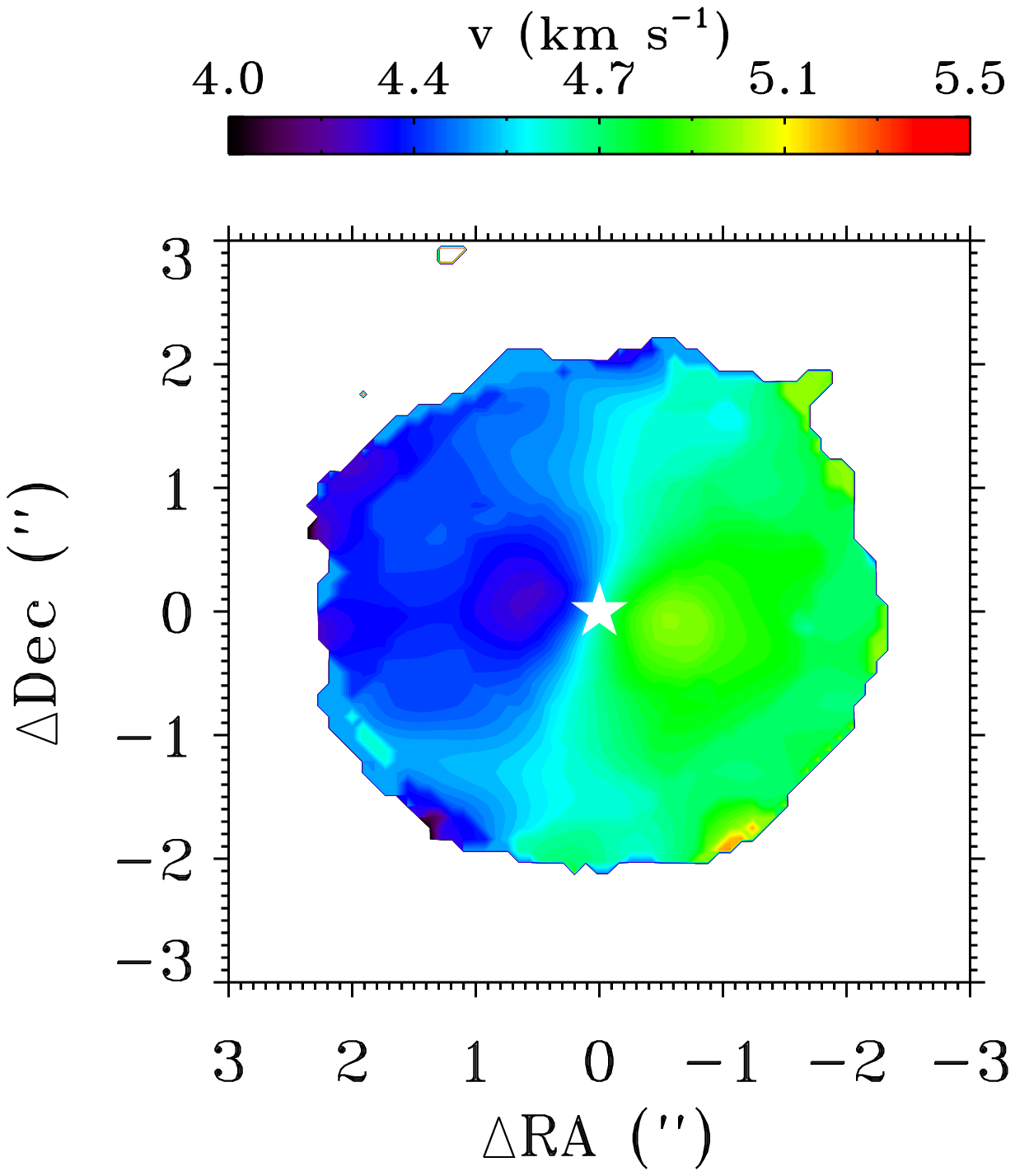}}\\
\caption{$^{12}$CO channel maps for each observed target (continued)}
\label{fig:channelmap2}
\end{figure*}

\end{document}